\documentclass[draft]{agujournal2018}
\usepackage{apacite}
\usepackage{amsmath}

\usepackage{url} 
\usepackage{textcomp}
\draftfalse

\journalname{JGR-Atmospheres}

\begin{document}

%
%


\title{The Ion and Charged Aerosol Growth Enhancement (ION-CAGE) code: A numerical model for the growth of charged and neutral aerosols}

%
%




\authors{J. Svensmark\affil{1,2}, N. J. Shaviv\affil{3}, M. B. Enghoff \affil{2}, H. Svensmark\affil{2}}


\affiliation{1}{DARK, Niels Bohr Institute, University of Copenhagen, Lyngbyvej 2, 4. sal, 2100 Copenhagen, Denmark}
\affiliation{2}{National Space Institute, Danish Technical University, Elektrovej 327, 2800 Kgs. Lyngby, Denmark}
\affiliation{3}{Racah Institute of Physics, Hebrew University of Jerusalem, Giv\textquotesingle at Ram, Jerusalem 91904, Israel}





\correspondingauthor{Jacob Svensmark}{jacob.svensmark@nbi.ku.dk}




\begin{keypoints}
\item A numerical aerosol growth model is developed to consider growth of charged and neutral aerosols
\item The model includes terms attributed to the condensation of small ion clusters onto aerosols
\item Modeled growth rates are enhanced by ion condensation in agreement with recent experiments
\end{keypoints}

%
%


\begin{abstract}
The presence of small ions influences the growth dynamics of a size distribution of aerosols. Specifically the often neglected mass of small ions influences the aerosol growth rate, which may be important for terrestrial cloud formation. To this end, we develop a numerical model to calculate the growth of a species of aerosols in the presence of charge, which explicitly includes terms for ion-condensation. It is shown that a positive contribution to aerosol growth rate is obtained by increasing the ion-pair concentration through this effect, consistent with recent experimental findings. The ion-condensation effect is then compared to aerosol growth from charged aerosol coagulation, which is seen to be independent of ion-pair concentration. The model source code is made available through a public repository.
\end{abstract}
\section{Introduction}
There is no shortage of numerical models aimed at describing the production, growth and transport of aerosols for a broad range of parameters, environments and applications. Often the scenarios that these models describe turn out very demanding from a computational point of view. Thus, as with any model, a number of optimizations and compromises must be made in terms of included effects and mechanisms, dimensionality, resolution, etc. Recent advances in our understanding of the interplay between atmospheric aerosols and electrical charge has heightened the need for aerosol growth modeling that takes charge into account in a detailed way, as charge seems to act as an enhancing agent for aerosol growth rate under atmospheric conditions relevant for marine aerosols \citep[hereafter "SESS17 paper"]{Svensmark2017}. In order to conduct tests of theoretical assumptions, or design and understand future experimental efforts in which charged aerosols are present, a complementary simple, modifiable, open-source numerical model can be helpful. Several advanced models describing aerosol dynamics have been developed, e.g., \cite{pierce2009,yu2001,laakso2002,kazil2004,leppa2009,mcgrath2012}. While some of these models take the charge of aerosols into account, none so far implement the mechanism of ion-induced condensation, i.e. the accelerated growth caused by the mass of ions. Others, such as \cite{prakash2003} are open-source, however do not include charge at all.

In the present work we introduce the Ion and Charged Aerosol Growth Enhancement (ION-CAGE) model: A zero-dimensional box model capable of solving the general dynamics equation (GDE) numerically for both single-charged and neutral aerosol species. Uniquely this model includes terms considering the addition of the non-zero mass of ions upon interactions with aerosols. It is coded in a way that allows for the easy in- or exclusion of a selection of the usual GDE terms, as well as code structure that can easily be modified to include new terms where desired. For this work the model is set up to simulate the growth of sulfuric acid aerosols in an environment corresponding to clean marine air, as relevant for the SESS17 experiment. The numerical solutions involving ion accelerated growth of aerosols demonstrate excellent agreement with the SESS17 theoretical results. To further test the theory, it would be useful to incorporate it into more advanced models, like those mentioned above. However, the results presented here already makes it feasible to test the SESS17 theory in computationally heavy global circulation models. 

In the following sections we present the terms of the GDE that are included in the model and demonstrate that it operates in correspondence with expectations. We apply the model by reproducing and expanding on the results of the SESS17 paper, by probing the growth rate of aerosols exposed to varying concentrations of ions-pairs and aerosols, considering also aerosol coagulation. 

\section{Model overview}
The ION-CAGE model considers the temporal evolution of number concentrations for the following species: Neutral and charged small stable and condensable molecular clusters (henceforth "neutral clusters" and "ions" respectively) and neutral as well as singly charged aerosols of both signs. Ions and neutral clusters are set at a certain diameter and density, whereas the aerosols exist at a number $N_{\mathrm{tot}}$ of logarithmically spaced nodes in volume as indicated by the index $k\in [1,N_{\mathrm{tot}}]$. This helps the model span the several orders of magnitude in volume that corresponds to aerosol diameters of typically between $1\,$nm and  $100\,$nm. Given a set of initial conditions for a distribution of neutral clusters, ions, and aerosols along all volume nodes, as well as interaction coefficients, production and loss parameters and other relevant inputs (see below for all parameters), the initial distribution is propagated through forward integration of the GDE using a 4th order Runge-Kutta algorithm. As shall be explored below, the GDE contains terms describing the nucleation of new aerosols, the condensation of neutral clusters and ions onto the aerosols, the coagulation of two aerosols into a single larger aerosol and production and loss terms all while keeping track of the exchange of charge between all species, except between neutral clusters and ions. Importantly the mass of the ion can be taken into account such that its small but important contribution to the condensation term is included.  The model is written in the FORTRAN language, and its source code is downloadable from a public repository at \url{https://github.com/jacobsvensmark/ioncage}.

\subsection{Size nodes}\label{Sec:nodes}
To create a set of $N_{\mathrm{tot}}$ volume nodes $v_k$ distributed linearly in logarithmic space between two volumes $v_{\mathrm{min}}$ and $v_{\mathrm{max}}$, we use the following 
\begin{equation}
v_k = 10^{\log \left( v_{\mathrm{min}}\right) + \frac{k-1}{N_{\mathrm{tot}}-1} \left(\log \left( v_{\mathrm{max}}\right) - \log \left(v_{\mathrm{min}}\right)\right)},
\end{equation}
where $k$ is an integer between 1 and $N_{\mathrm{tot}}$. In this way two neighboring nodes are separated by a factor $C$ such that $v_k = Cv_{k-1}$. 
\begin{figure}[ht!]
   \centering
\includegraphics[width=12cm]{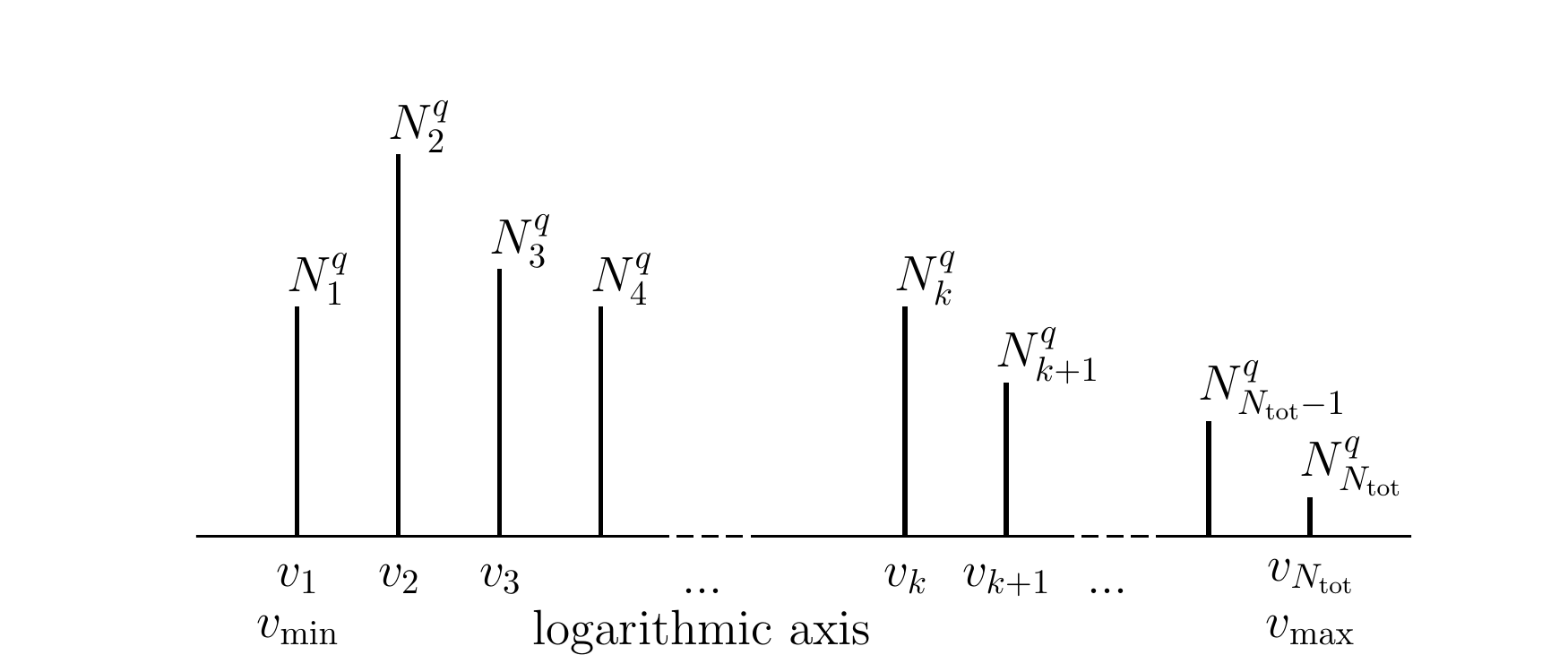}
   \caption{Overview of the node structure of the volume bins, and the concentrations at each node for a given charge $q$. Note that the scale is logarithmic, that each node is a factor of $C$ larger than its smaller neighbor: $v_k = Cv_{k-1}$.}
   \label{fig:nodes}
\end{figure}
Aerosols in the model have a charge of $q=[0,-.+]$, i.e., 0, -1 or 1 elementary charge. The concentration of aerosols at volume $v_k$ and charge $q$ is denoted $N_k^q$, and thus the model keeps track of $3N_{\mathrm{tot}}$ aerosol concentrations. A diagram of the node structure can be seen in Figure \ref{fig:nodes}. Additionally the model has neutral condensable clusters at concentration $n^0$, as well as condensable positive and negative ions at concentrations $n^+$ and $n^-$, all of which we refer to as monomers.
These concentrations constitute the state of the model, and following a set of terms that make up the GDE of this model, the state is propagated forward in time. The terms that govern the dynamics of this process are presented in the next section.

\section{Interactions}
In the following we formulate the interactions  governing the temporal derivatives between charged and neutral aerosols, ions and neutral condensable clusters.

\begin{figure}[ht!]
   \centering
\includegraphics[width=14cm]{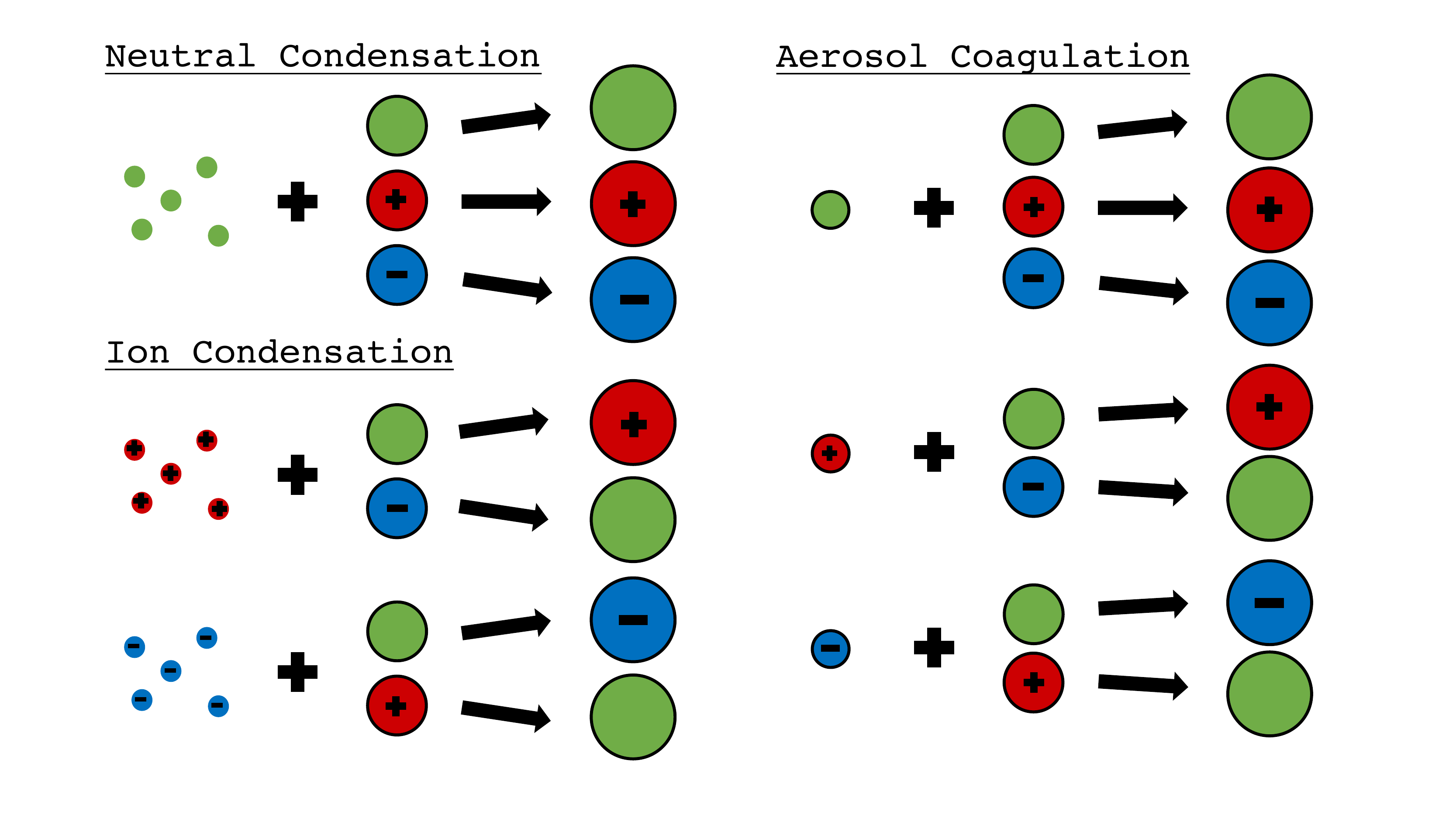}\\
\vspace{-0.5cm}
\includegraphics[width=14cm]{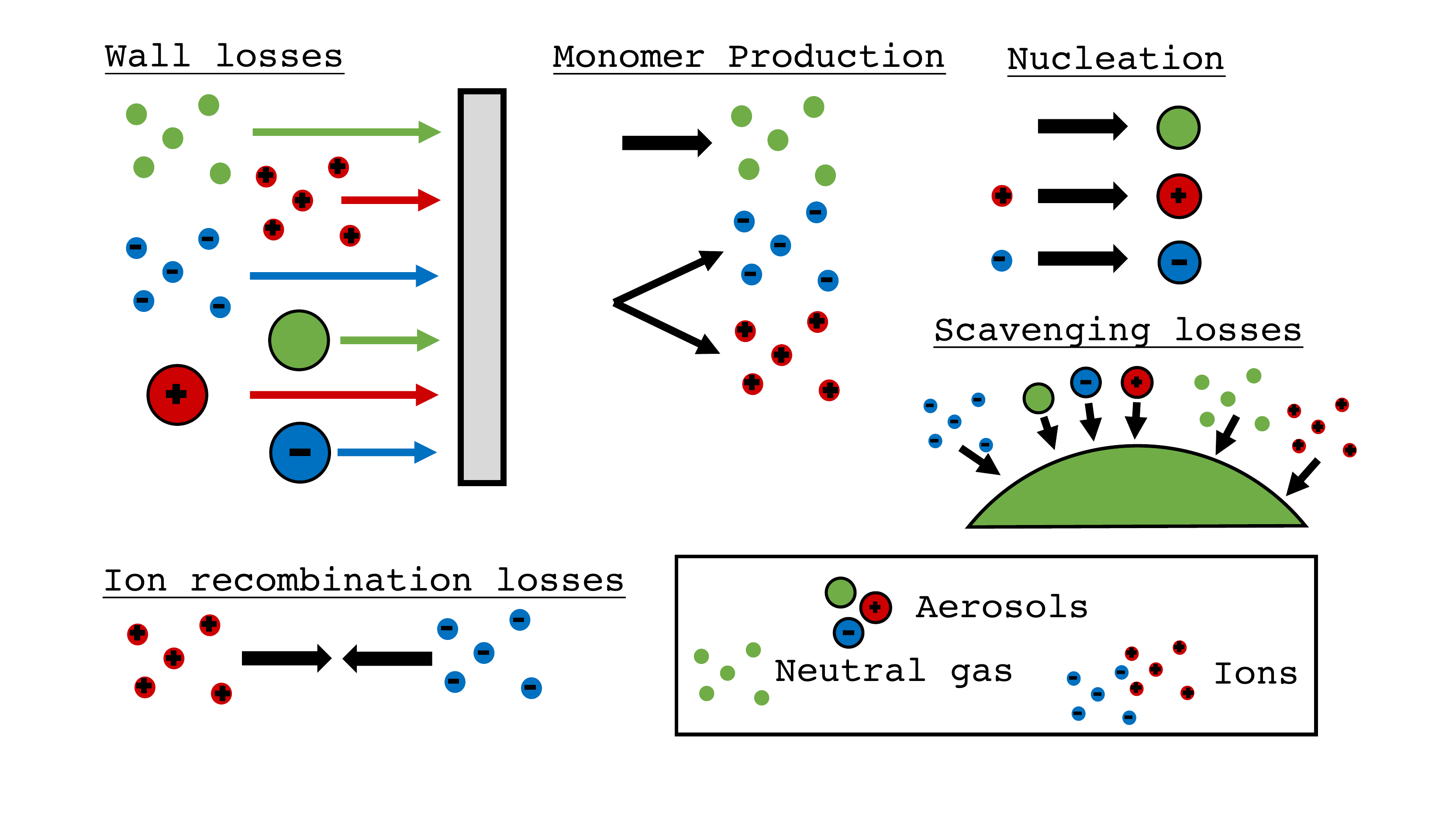}\\
\vspace{-1cm}
   \caption{Diagramatic overview of the interactions included in the model. The interacting species amount to condensing neutral clusters and positive and negative clusters i.e. ions depicted as small groups of molecules, as well as neutral, positive and negative aerosols, in a wide range of diameters. Note in particular the terms depicted in the lower parts of the diagram, which shows how volume is removed or introduced to the model.}
   \label{fig:diagram}
\end{figure}

A diagram of the mechanisms included by the model can be seen in Figure \ref{fig:diagram}. Since the sizes of the aerosols of interest are typically in the range 1 - 100$\,$nm, only singly charged aerosols are considered. In depth discussions on the information presented here can be found in e.g. \citet{seigneur1986,zhan1999,seinfeldpandis}.

The overall dynamical equations for aerosols at charge $q=[0,-,+]$ are
\begin{equation}\label{eq:dyneq_aerosol}
\frac{dN_k^q}{dt} = \left. \frac{d N_k^q}{d t}\right] _{\mathrm{nucleation}} 
                           +  \left. \frac{d N_k^q}{d t}\right] _{\mathrm{condensation}} 
                           +  \left. \frac{d N_k^q}{d t}\right] _{\mathrm{coagulation}} 
                           +  \left. \frac{d N_k^q}{d t}\right] _{\mathrm{loss}}.
\end{equation}
The first term on the right hand side of Equation \ref{eq:dyneq_aerosol} contains a simple contribution to the nucleation of new aerosol particles of charge $q=[0,-,+]$. The second contains the condensation of neutral clusters and ion mass onto aerosols, while keeping track of the charge of the aerosols in this process. The third term is the most computationally expensive term and describes the coagulation of two aerosols into a larger common aerosol, again taking charge into account. The fourth and final term adds losses in the form of scavenging by large particles and losses to walls required for modelling an experimental situation.

Similarly for the neutral cluster and ion concentrations $n^q$ the terms of the overall GDE are
\begin{equation}
\frac{dn^q}{dt} = \left. \frac{d n^q}{d t}\right] _{\mathrm{nucleation}} 
                       +  \left. \frac{d n^q}{d t}\right] _{\mathrm{condensation}} 
                       +  \left. \frac{d n^q}{d t}\right] _{\mathrm{production}} 
                       +  \left. \frac{d n^q}{d t}\right] _{\mathrm{loss}}.
\end{equation}
Here the first and second term on the right hand side describe the same processes as those of Equation \ref{eq:dyneq_aerosol}. The third term includes production of monomers, and the final loss term takes ion recombination into account in addition to wall losses. The number of coupled equations become $3N_{\mathrm{tot}}$ +3, where the numbers three denote the three charging states of the aerosols and three rate equations for neutral clusters and ions. The code is written in a modular form to allow easy modification, expansion or decoupling of each of the terms. 

\subsection{Nucleation}
A nucleation term specifies the amount of new aerosols generated per unit time at a specified critical volume $v^q_{\ast}$, where $q=[0,-,+]$ denotes the charge. The production of new aerosols at this size is set to a fixed rate $J^q$. The volume of the nucleated particles may fall in between two nodes, or below the lowest node. Following the design of \citep{prakash2003} the number concentration is then scaled by volume and placed in the node immediately above it, such that 
\begin{equation}\label{eq:nucleation} 
\left. \frac{d N_k^q}{d t}\right] _{\mathrm{nucleation}} =
\begin{cases} 
J^q\frac{v^q_{\ast}}{v_1},\quad & \mathrm{for} \quad v^q_{\ast} \leq  v_1 \\
J^q\frac{v^q_{\ast}}{v_k},\quad & \mathrm{for} \quad v_{k-1} < v^q_{\ast} \leq  v_k \\
0,\quad & \mathrm{otherwise}.
\end{cases}
\end{equation}
Once ion-nucleation is applied it is important to maintain charge conservation. Positively and negatively nucleated aerosols take their charge from the ions in a number corresponding to what was nucleated at the critical volume, and this charge has to be subtracted from the positive and negative ion equations, i.e.,\
\begin{eqnarray}
 \left. \frac{d n^-}{d t}\right] _{\mathrm{nucleation}} & = & -J^-, \nonumber \\ 
 \left. \frac{d n^+}{d t}\right] _{\mathrm{nucleation}} & = & -J^+.
\end{eqnarray}
This simplistic view on aerosol nucleation is easily expanded. The nucleation could be modified to depend on other quantities in the model such as neutral cluster and ion concentrations, temperature and/or further parametrisation \citep{yu2010,dunne2016,Maattanen2017}.

\begin{table}
\caption{Table with variables and their descriptions used in the present paper.}
\centering
\begin{tabular}{llp{110mm}}
\hline
Symbol  & Unit & Description \\
\hline

$v_k$ & m$^3$ & Volume of aerosol node $k$.\\
$v^q$ & m$^3$ & Volume of monomer of charge $q$.\\
$v_{min}$ & m$^3$ & Volume of the smallest node.\\
$v_{max}$ & m$^3$ & Volume of the largest node.\\
$v^q_{\ast}$ & m$^3$ & Critical volume at which nucleation tales place for aerosols at charge $q$.\\
$t$ & s & Time.\\
$C$	&	&	Ratio between volumes of two neighbouring nodes\\
$N_{tot}$ & & Number of logarithmically spaced size nodes.\\
$q,p$  &  & Indices specifying charge, may be $\left(0,-,+ \right)$ dictating negative, neutral and positive charge respectively.\\
$i,j,k$ &  & Indices specifying node number.\\
$\sigma$ & & Index with values {1,2,3,4}.\\
$N^q_k$ & m$^{-3}$ & Particle concentration of particles at node number $k$ and charge $q$. \\
$N_L$ &  m$^{-3}$ & Concentration of large mode aerosols used in scavenging loss term. \\
$n^q$ & m$^{-3}$ & Concentrations of neutral ($q=0$) or ion monomers ($q=-$ or $q=+$).\\ 
$J^q$ & m$^{-3}$s$^{-1}$ & Nucleation rate of charged or neutral aerosols.\\
$\beta_k^{qp}$ & m$^3$s$^{-1}$ & Condensation coefficient of interactions between a monomer of charge $q$ and an aerosol of charge $p$ at size node $k$.\\
$\kappa_{i,j}^{qp}$ & m$^3$s$^{-1}$ & Coagulation coefficient of interactions between an aerosol from size node $i$ and charge $q$, and another aerosol from size node $j$ and charge $p$.\\
$\beta _ L ^{q0}$ &  m$^3$s$^{-1}$ & Condensation coefficient for loss of monomers of charge $q$ onto large mode scavenging aerosols.\\
$\kappa_{k,L}^{q0}$ & m$^3$s$^{-1}$ & Coagulation coefficient for losses of an aerosol from size node $k$ and charge $q$, and a large mode scavenging aerosol.\\
$S_{i,j}$ & & Splitting matrix denoting the fractional contribution of two coagulating particles from node $i$ and $j$ to two neighboring nodes.\\
$P$ & m$^{-3}$s$^{-1}$ & Neutral cluster production rate.\\
$Q$ & m$^{-3}$s$^{-1}$ & Ion-pair production rate.\\
$\alpha$ & m$^3$s$^{-1}$ & Recombination coefficient.\\
\hline
\end{tabular}
\label{tab:variables}
\end{table}

\subsection{Condensation}\label{sec:cond}
Three channels of condensation of clusters onto aerosol particles are allowed: Neutral clusters of concentration $n^0$ condensing onto all aerosols, and two classes of ions $n^+$ and $n^-$ with a single charge condensing onto neutral aerosols or aerosols of opposite sign. Schematically we write:
\begin{equation}
\label{Eq:condensation1}
\left. \frac{d N_k^q}{d t}\right] _{\mathrm{condensation}} = \left. \frac{d N_k^q}{d t}\right] _{\mathrm{cond},n^0} + \left. \frac{d N_k^q}{d t}\right] _{\mathrm{cond},n^+} +  \left. \frac{d N_k^q}{d t}\right] _{\mathrm{cond},n^-},
\end{equation}
where $q=[0,-,+]$ indicate the neutral, positive and negative charges respectively. This yields three equations of three terms summarized in the following equation:
\begin{equation}\label{Eq:1}
 \left. \frac{d N_k^q}{d t}\right] _{\mathrm{condensation}} = -\sum_p I^{qp}_k N_k^p +\sum_p I^{qp}_{k-1} N_{k-1}^p,
\end{equation}
where
\begin{eqnarray} 
 I^{qp}_k (r,t) = \left( \begin{array}{ccc} \label{eq:cond_matrix}
A_k^0 n^0 \beta_k^{00} & A_k^+n^+ \beta_k^{+-}& A_k^-n^- \beta_k^{-+}  \\
A_k^-n^- \beta_k^{-0} &  A_k^{0} n^{0} \beta_k^{0-} & 0 \\
A_k^+n^+ \beta_k^{+0} & 0 & A_k^{0} n^{0} \beta_k^{0+}\end{array} \right).
\end{eqnarray}
For an aerosol to grow from volume $v_k$ to $v_{k+1}$, a number $(v_{k+1}-v_{k})/v^q$ of condensing monomers of volume $v^q$ is needed. The coefficients $A^q_k=v^q/(v_{k+1}-v_{k})$ are thus the fractional volume that one condensing monomer contributes to growing an aerosol from volume $v_k$ to $v_{k+1}$. This is analogous to the treatment of condensation in e.g. \citet{prakash2003}. $q$ and $p$ indicate the charge, and the $\beta_k^{qp}$ denotes the interaction coefficient between a monomer (neutral or ion) of charge $q=[0,-,+]$ and an aerosol at size node $k$ and charge $p=[0,-,+]$. The diagonal elements account for the usual condensation of neutral clusters, while the other four non-zero elements make up the terms related to the condensation of ions. The 0-terms in the matrix represent the negligible interactions between like-charged ions and aerosols.

As monomers condense onto the aerosols, neutral cluster and ion concentrations change accordingly:
\begin{eqnarray}
\label{Eq:condensation2}
\left. \frac{d n^0}{d t}\right] _{\mathrm{condensation}} &=& - n^0\sum_{k=1}^{N_{\mathrm{tot}}} \left(  \beta_k^{00}N_k^{0} +\beta_k^{0+}N_k^{+} +\beta_k^{0-}N_k^{-} \right), \\ \nonumber
\left. \frac{d n^-}{d t}\right] _{\mathrm{condensation}} &=& - n^- \sum_{k=1} ^{N_\mathrm{tot}}\left(  \beta_k^{-0} N^0_k + \beta_k^{-+} N^+_k \right), \\ \nonumber
\left. \frac{d n^+}{d t}\right] _{\mathrm{condensation}} &=& - n^+ \sum_{k=1} ^{N_\mathrm{tot}}\left( \beta_k^{+0}N^0_k  + \beta_k^{+-} N^-_k \right).
\end{eqnarray}
The code can easily switch on/off the terms related to the usual condensation and ion-condensation separately. 
The reverse process, loss of aerosol volume due to evaporation, is not implemented in the present model since only stable clusters are considered. Note that evaporation is not needed for demonstrating the novel feature of ion-condensation. It is straight forward to include evaporation as a future extension of the model. 
\subsection{Monomer production}
A term describing the production of the monomers is included in the model in the following simple way
\begin{eqnarray}
\left. \frac{d n^0}{d t}\right] _{\mathrm{production}}  &=& P, \\
\left. \frac{d n^-}{d t}\right] _{\mathrm{production}}  &=& Q, \\
\left. \frac{d n^+}{d t}\right] _{\mathrm{production}}  &=& Q. 
\end{eqnarray}
Here, $P$ is the production rate of the neutral cluster elements $n^0$, $Q$ is the ion-pair production rate. Both rates are input to the model as constants, but can easily be generalized to describe more advanced cases. The code allows for bypassing the production rates to keep a constant concentration of any of the $n^q$.

\subsection{Coagulation}
When particles of volume $v_i$ and $v_j$ coagulate, they produce a new particle with a volume $v_i+v_j$, which may in general lie between two nodes $k$ and $k+1$. Thus the coagulated volume needs to be split and distributed between $N_k^{q}$ and $N_{k+1}^{q}$. To do so, we calculate a matrix giving for each $i,j$ pair the index $k$ of the node immediately below the new coagulated volume:
\begin{equation}
V_{ij} = f_k(v_i + v_j) = \mathrm{floor} \left[  N_{\mathrm{tot}} \left( \frac{\log (v_i + v_j)  - \log v_{\mathrm{min}}}{\log v_{\mathrm{max}} - \log v_{\mathrm{min}} } \right) \right] .
\end{equation}
$v_{\mathrm{min}}$ is the volume of the smallest node, and $v_{\mathrm{max}}$ that of the largest. We also calculate a splitting fraction function
\begin{equation}
f_s(v) = \frac{ v_{k+1} - v }{ v_{k+1} - v_k},
\end{equation}
where $f_s(v)$ is a function giving the volume fraction added to node $k=f_k(v)$ while $1-f_s(v)$ is added to node $k+1$. The function $f_s(v)$ is defined such that $v = f_s(v_k) v_k + (1-f_s(v_k) ) v_{k+1}$, implying that the total volume is conserved with this definition of splitting. 

To save computing time, a splitting matrix is calculated during initialization and used throughout the calculation. Its terms are
\begin{equation}
S_{ij} = f_s (v_i + v_j) = \frac{ v_{k+1} - (v_i + v_j) }{ v_{k+1} - v_k}.
\end{equation}

The charge gives rise to four types of coagulation between the aerosols: Neutral with neutral,  positive with neutral, neutral with negative, and positive with negative. These have four corresponding coagulation coefficient matrices: $\kappa_{ij}^{00}, \kappa_{ij}^{+0}, \kappa_{ij}^{0-}, \kappa_{ij}^{+-}$, each of dimension $N_{\mathrm{tot}} \times N_{\mathrm{tot}}$. Coagulation of like-signed aerosols is neglected, as the model only incorporates single-charge species.
During one time step, we can then calculate the interaction terms from each coagulation contribution:
\begin{equation}
\widetilde{I^{\sigma}_{ij}}=
\begin{pmatrix}
I_{ij}^{00} \\
I_{ij}^{+0} \\
I_{ij}^{0-} \\
I_{ij}^{+-} 
\end{pmatrix}
=
\begin{pmatrix}\label{eq:coag_interaction_terms}
\kappa_{ij}^{00} N_i^0 N_j^0  \\
\kappa_{ij}^{+0} N_i^+ N_j^0 \\
\kappa_{ij}^{0-} N_i^0 N_j^-   \\                    
\kappa_{ij}^{+-} N_i^+ N_j^-   
\end{pmatrix}.
\end{equation}
The rules for mapping coagulation of two particles of given charges into its electrically relevant node are summarized by the following three matrices:
\begin{equation}\label{eq:coag_matrix}
K^{q\sigma}=
\begin{pmatrix}
\frac{1}{2}    &     0     &     0     &     1\\
0                   &     0     &     1     &     0\\
0                   &     1     &     0     &     0
\end{pmatrix},\quad
L^{q\sigma}=
\begin{pmatrix}
-\frac{1}{2}   &     0    &     -1     &     0\\
0                   &     0     &     0     &     0\\
0                   &    -1     &     0     &     -1
\end{pmatrix}, \quad
M^{q\sigma}=
\begin{pmatrix}
-\frac{1}{2}   &    -1     &     0     &     0\\
0                   &     0     &    -1     &    -1\\
0                   &    0      &     0     &     0
\end{pmatrix}. 
\end{equation}
As before, $q$ denotes the charge of the end-product, and $\sigma$ is a number between 1 and 4 denoting one of the four possible types of coagulation.

As aerosols at size node $i$ coagulate with those at node $j$, and distribute their joined volume between node $k$ and $k+1$ where $k$ is a function of $i$ and $j$, the rate of change in the four nodes is as follows:
\begin{eqnarray}
\left.\frac{dN_k^{q}}{dt}        \right] _{\mathrm{coag}} &=& S_{ij}\sum_{\sigma} K^{q\sigma} \widetilde{I^{\sigma}_{ij}},\nonumber\\
\left.\frac{dN_{k+1}^{q}}{dt} \right] _{\mathrm{coag}} &=& (1-S_{ij})\sum_{\sigma} K^{q\sigma} \widetilde{I^{\sigma}_{ij}},\nonumber\\
\left.\frac{dN_{i}^{q}}{dt}      \right] _{\mathrm{coag}} &=& \sum _{\sigma} L^{q\sigma}\widetilde{I^{\sigma}_{ij}},\nonumber \\
\left.\frac{dN_{j}^{q}}{dt}      \right] _{\mathrm{coag}} &=& \sum _{\sigma} M^{q\sigma}\widetilde{I^{\sigma}_{ij}}.
\end{eqnarray}
Repeating this for all combinations of $i$ and $j$ yields the total coagulation term. All coagulation channels are covered within the four terms of Eq (\ref{eq:coag_interaction_terms}) as the summation runs through all combinations of $i$ and $j$. The neutral - neutral interactions however are counted twice since e.g. the coagulation for $i=1$ and $j=2$ represents the same process as $i=2$ and $j=1$ if both particles are neutral. This is compensated for by the factor of $\frac{1}{2}$ on all neutral - neutral coagulation terms from the matrices of Eq. (\ref{eq:coag_matrix}).

\subsection{Loss terms}
When simulating an experimental situation where an atmospheric reaction chamber is used, loss of aerosols to the walls may be of importance. To accommodate this, a size dependent loss on $N_k^q$, $n^{\pm}$, and $n_{0}$ is added to the model. An example of such a loss term was found empirically to depend on aerosol radius $a$ as $dN_k/dt = -\lambda(a/a_0)^{-\gamma}N_k$, where $\gamma = 0.69$ and $\lambda=6.2\times 10^{-4}\,$s$^{-1}$ and $a_0=1\,$nm, for an 8$\,$m$^3$ reaction chamber \citep{svensmark2013}. Furthermore, we include a losses to large aerosols at concentration $N_L$ and diameter $d_{L}$. Extending this to charged aerosols and monomers, the wall loss term can be written for the aerosols as
\begin{equation}
\left. \frac{d N_k^q}{d t}\right] _{\mathrm{loss}} = -\lambda\left( \frac{d_k}{d_\mathrm{loss}}\right) ^{-\gamma}N_k^q - N_k^q\kappa_{kL}^{q0}N_L,
\end{equation}
where $d_\mathrm{loss}=2\,$nm and $d_k$ is the diameter corresponding to volume node $k$, and $\kappa_{kL}^{q0}$ is the coagulation coefficient between the $k$'th aerosol of charge $q$ and the large aerosol. For the monomers we have  
\begin{equation}\label{eq:loss_wall_mon}
\left. \frac{d n^q}{d t}\right] _{\mathrm{loss}} = -\lambda \left(\frac{d^q}{d_\mathrm{loss}}\right) ^{-\gamma}n^q - n^q\beta_L^{q0}N_L,
\end{equation}
where $d^q$ are the diameters of the monomers, and $\beta_L^{q0}$ is the interaction coefficient between the $k$'th aerosol of charge $q$ and the large aerosol. Here $\lambda$, $\gamma$, $d_\mathrm{loss}$, $d_L$ and $N_L$ as well as the coefficents serve as inputs to the model. Further losses due to ion-ion recombination are implemented through the recombination coefficient $\alpha$, such that in addition to Equation \ref{eq:loss_wall_mon} we have
\begin{eqnarray}
\left. \frac{d n^-}{d t} \right] _{\mathrm{loss}} &=& - \alpha n^-n^+,\\
\left. \frac{d n^+}{d t} \right] _{\mathrm{loss}} &=& - \alpha n^-n^+.
\end{eqnarray}
This means that recombining ions effectively disappear from the model. Note that the recombination coefficient $\alpha=1.6\times 10^{-6}\,$cm$^3$s$^{-1}$ is used, and set as a constraint on the coefficients $\beta_k^{qp}$ and $\kappa_{ij}^{qp}$ we use for demonstration purposes here (see section on parameters and the Appendix). Other types of loss such as rain out could be added in the code. Also condensation of monomers onto aerosols in the largest size node, or coagulation out of the range spanned by the size nodes constitutes losses from the model.

\subsection{Parameters}\label{sec:parameters}
To accurately simulate the interactions of aerosols through the channels mentioned above, the interaction coefficients $\beta_k^{qp}$ and $\kappa_{ij}^{qp}$ must be described as accurately as possible. These interactions depend on the species that are modeled and the environment they exist in. We model the interaction caused by condensation of hydrated sulphuric acid clusters onto sulphuric acid-water-aerosols and ions in later parts of the present work. To do that we model neutral clusters, ions and aerosols as electrostatically interacting spheres including their image induction in one another. Furthermore Van der Waals forces, viscous forces and dipole moment of the sulfuric acid population are included along with Brownian coagulation to make up the coefficients $\beta$. Further information on how the coefficients have been calculated can be found in the Appendix. Figure \ref{fig:beta_II} displays the $\beta$ coefficients as a function of aerosol diameter. The choice of parameters for this figure is summarized in Table \ref{tab:parameters_SESS}, and chosen to model the behavior of hydrated sulphuric acid clusters, ions, and aerosols at 300$\,$K. These coefficients are used throughout the paper, unless otherwise is explicitly stated. The model takes condensation and coagulation coefficient tables as input prior to any calculations, and thus any relevant coefficients may be provided.

\begin{figure}[ht]
   \centering
\includegraphics[width=13cm]{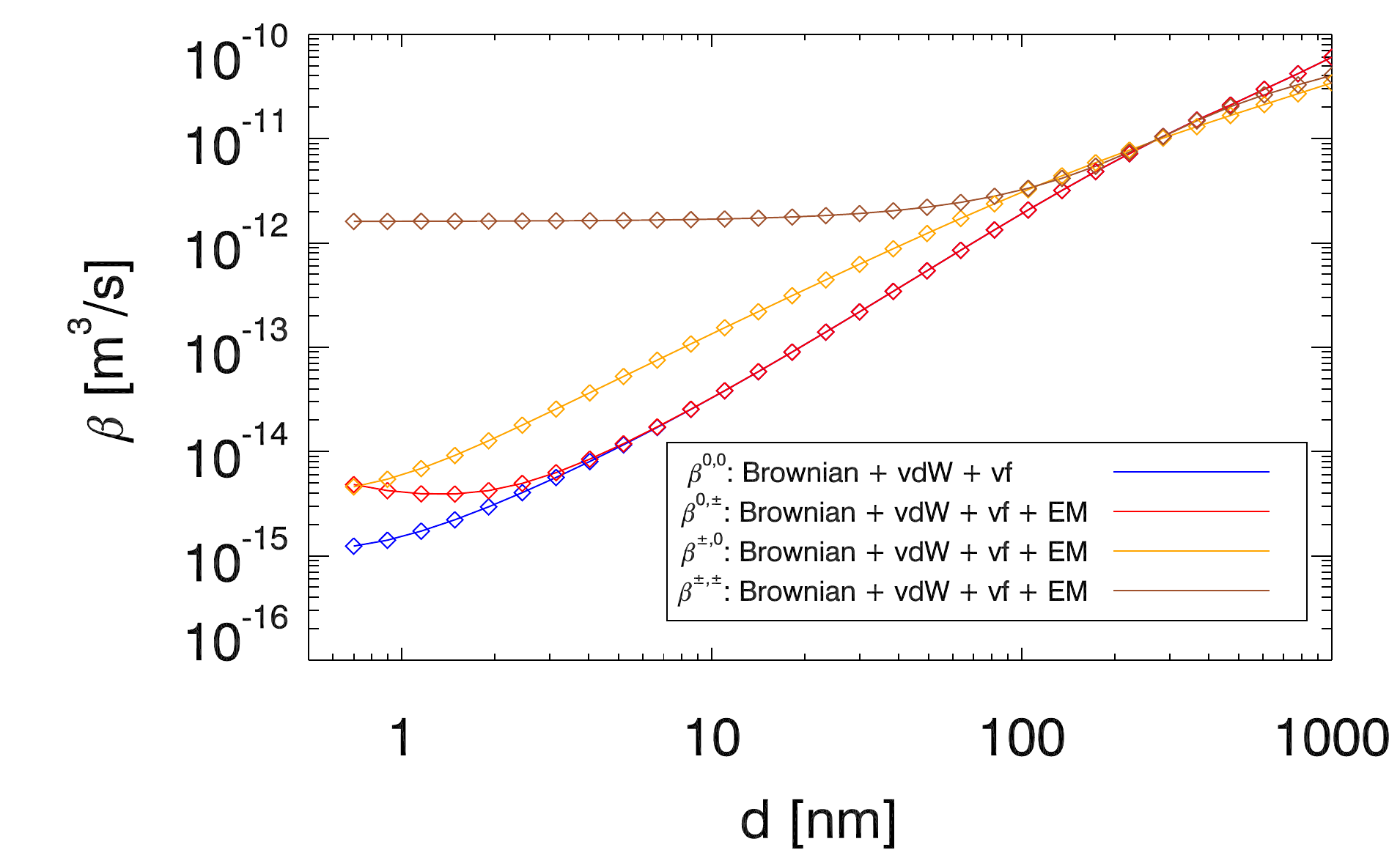}
   \caption{The condensation coefficients $\beta$ for a neutral cluster monomer or charged ion interacting with aerosols. The potentials are: vdW: Van der Waals potential.  vf: Viscious forces. EM: Electrostatic potential. The neutral clusters are here H$_2$SO$_4$ and are assumed to have a mass of $100\,$AMU. Ions have masses of $225\,$AMU.}
   \label{fig:beta_II}
\end{figure}

\section{Benchmarking the model}
\begin{figure}[ht!]
   \centering
   \vspace{-2cm}
\includegraphics[width=8cm]{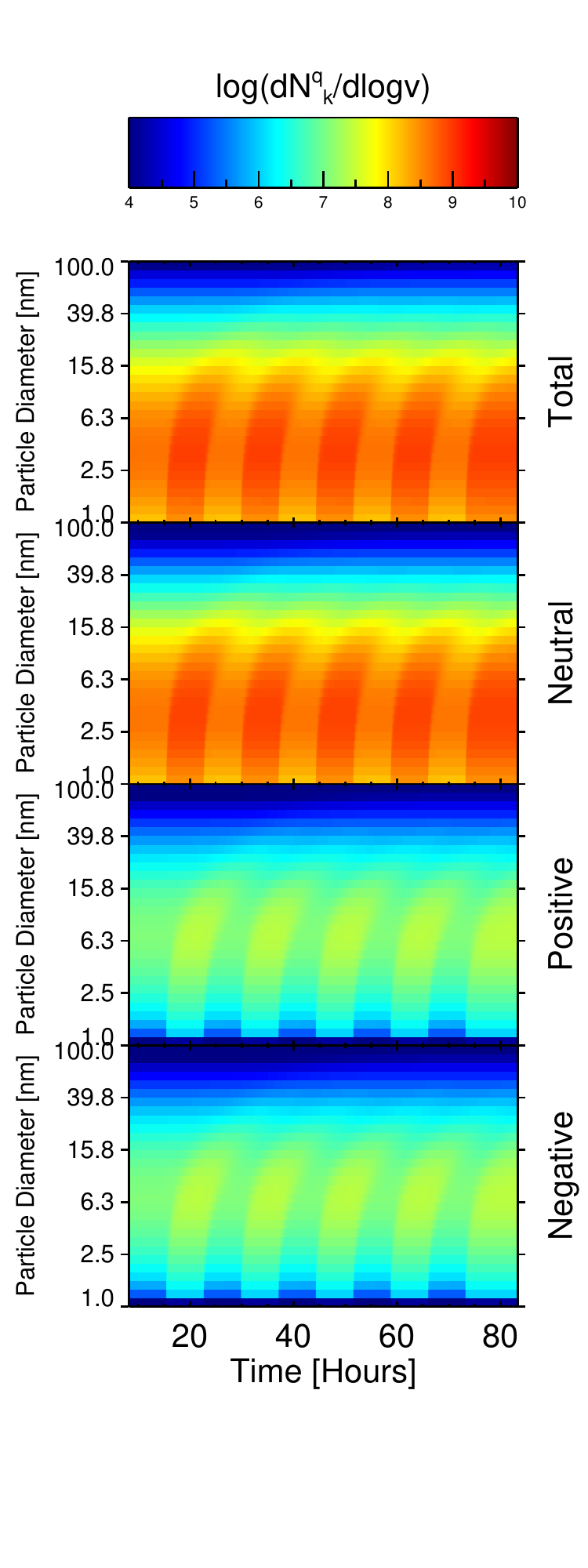}
   \caption{16 hour cycles of $J^0=0.1\,$cm$^{-3}$s$^{-1}$ and $J^0=0.2\,$cm$^{-3}$s$^{-1}$, with ionization rates also alternating between $Q=16\,$cm$^{-3}$s$^{-1}$ and $Q=500\,$cm$^{-3}$s$^{-1}$. Coagulation is not included for this particular calculation.}
   \label{fig:run}
\end{figure}

To validate the model, we test to see whether it can reproduce known solutions to the GDE for simple initial conditions. Thus a number of benchmark tests are performed.  First a visual representation of the model output is shown in Figure \ref{fig:run}. In this example the model was initiated with a steady state distribution. Then, after 8 hours the nucleation rate was increased by a factor of 2 along with a change in the ion-pair production rate from 16$\,$cm$^{-3}$s$^{-1}$ to 500$\,$cm$^{-3}$s$^{-1}$. After 8 additional hours the ion-pair production and nucleation rate was stepped back down, and this process was repeated 5 times. The figure displays in the first panel the distribution for all aerosols (i.e. $N^0_k+N^+_k+N^-_k$) as a function of time. In the lower panels are the neutral, positive and negative aerosols only also as a function of time. In all figures a clear growth profile is seen with a period of 16 hours.

\subsection{Testing condensation and coagulation}
\begin{figure}[htbp]
   \centering
       \includegraphics[width=12cm]{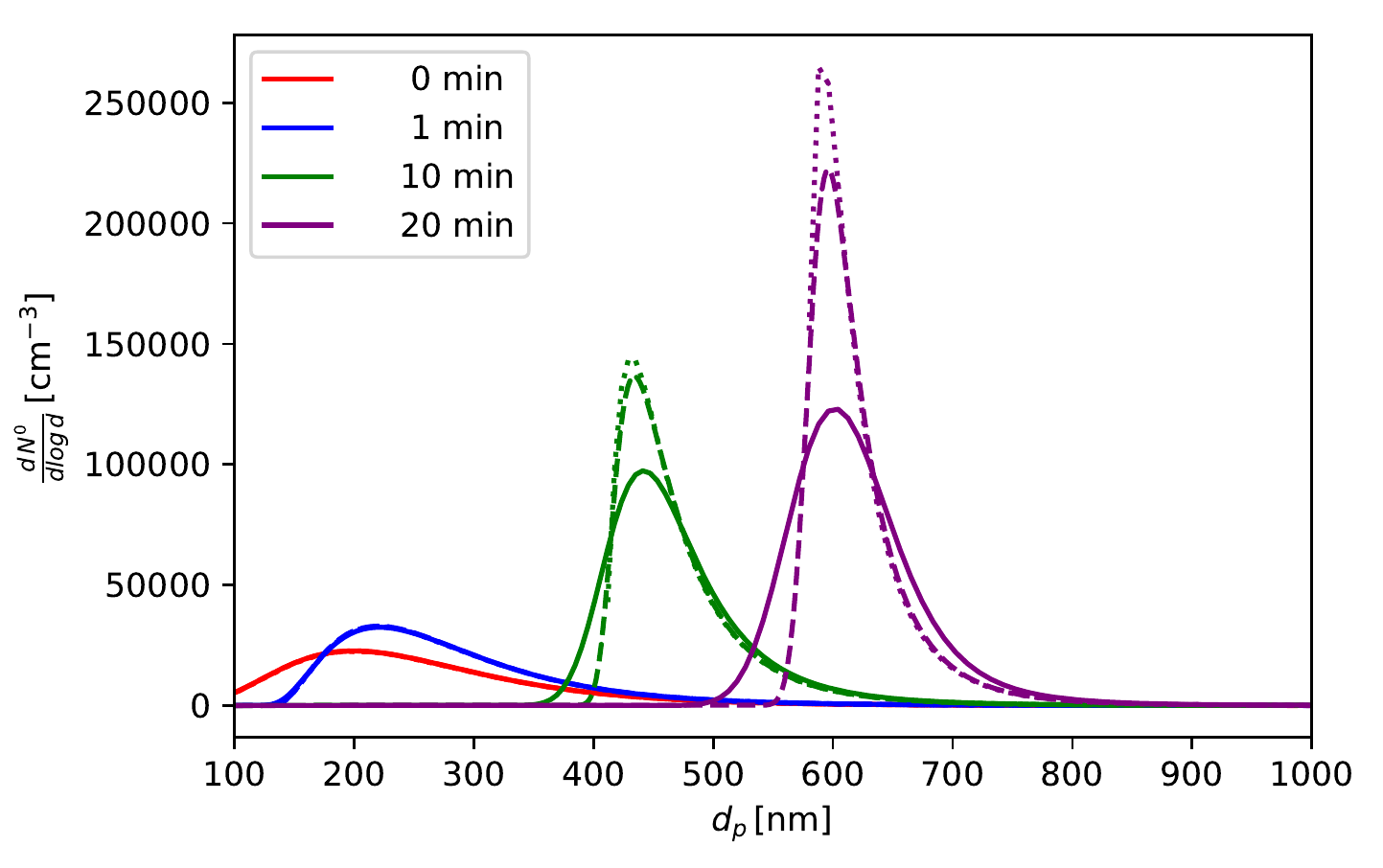}
       \caption{Pure condensation for an initial lognormal distribution of aerosols from equation \ref{eq:cond_initial}, using $\sigma_g=1.5$, $D_{g}=200\times 10^{-9}\,\mathrm{m}$ and $N_0=10000\,\mathrm{cm}^{-3}$. The condensation coefficients used for this reflect the choice of $(p_{i}-p_{\mathrm{i,eq}})=10^{-9}\,$atm, $M_i=100\,\mathrm{g/mol}$, $D_i=10^{-5}\,\mathrm{m}^2\mathrm{s}^{-1}$, $T=300\,$Km $\rho_p=1.2\,\mathrm{g}\,\mathrm{cm}^{-3}$). Dotted lines show the analytical solution, solid lines are snapshots from a coagulation-only simulation, and dashed lines are a similar simulation with a factor of higher density of volume nodes, lowering the numerical diffusion.}
    \label{fig:condensation_benchmark}
\end{figure}

To see that the condensation mechanism is on par with expectations, a known distribution of aerosols is propagated forward in time while neglecting all contributions to the GDE other than the condensation term. The numerical model output can then be compared to an analytic temporal evolution of the same initial conditions. The condensation equation can be written as 
\begin{equation}
\frac{\partial n(d,t)}{\partial t}+\frac{A}{d}\frac{\partial n(d,t)}{\partial d}=\frac{A}{d^2}n(d,t)
\end{equation}
where $A=4D_iM_i(p_i-p_{\mathrm{eq,i}})/RT\rho_p$ is a constant. An initial continuous lognormal aerosol distribution 
\begin{equation}\label{eq:cond_initial}
n(d,0)=\frac{N_0}{\sqrt{2\pi}d\,\ln{\sigma_g}}\exp\left( -\frac{\ln^2(d/D_g)}{2\ln^2\sigma_g}\right)
\end{equation}
is grown using only the coagulation section of the model (see eg. \citet{seinfeldpandis} Equation (13.26)). Here $N_0$ is the number of aerosols, while $D_g$ and $\sigma _g$ determine the shape of the distribution. Analytically, this solves to 
\begin{equation}
n(d,t)=\frac{d}{(d^2-2At)}\frac{N_0}{\sqrt{2\pi}d\,\ln{\sigma_g}}
\exp \left(-\frac{\ln^2\left[(d^2-2At)^{1/2}/D_g\right]}{2\ln^2\sigma_g}\right).
\end{equation}
In figure \ref{fig:condensation_benchmark}, the analytical solution as well as a corresponding simulation is seen for the parameters i.e. interaction coefficients written in the Figure caption. As can be seen, the peak of the distribution follows the analytical form as expected, however as the number of volume nodes decreases the numerical diffusion becomes more apparent.

\begin{figure}[htbp]
   \centering
   \hspace{-0.5cm}
        \includegraphics[width=7.6cm]{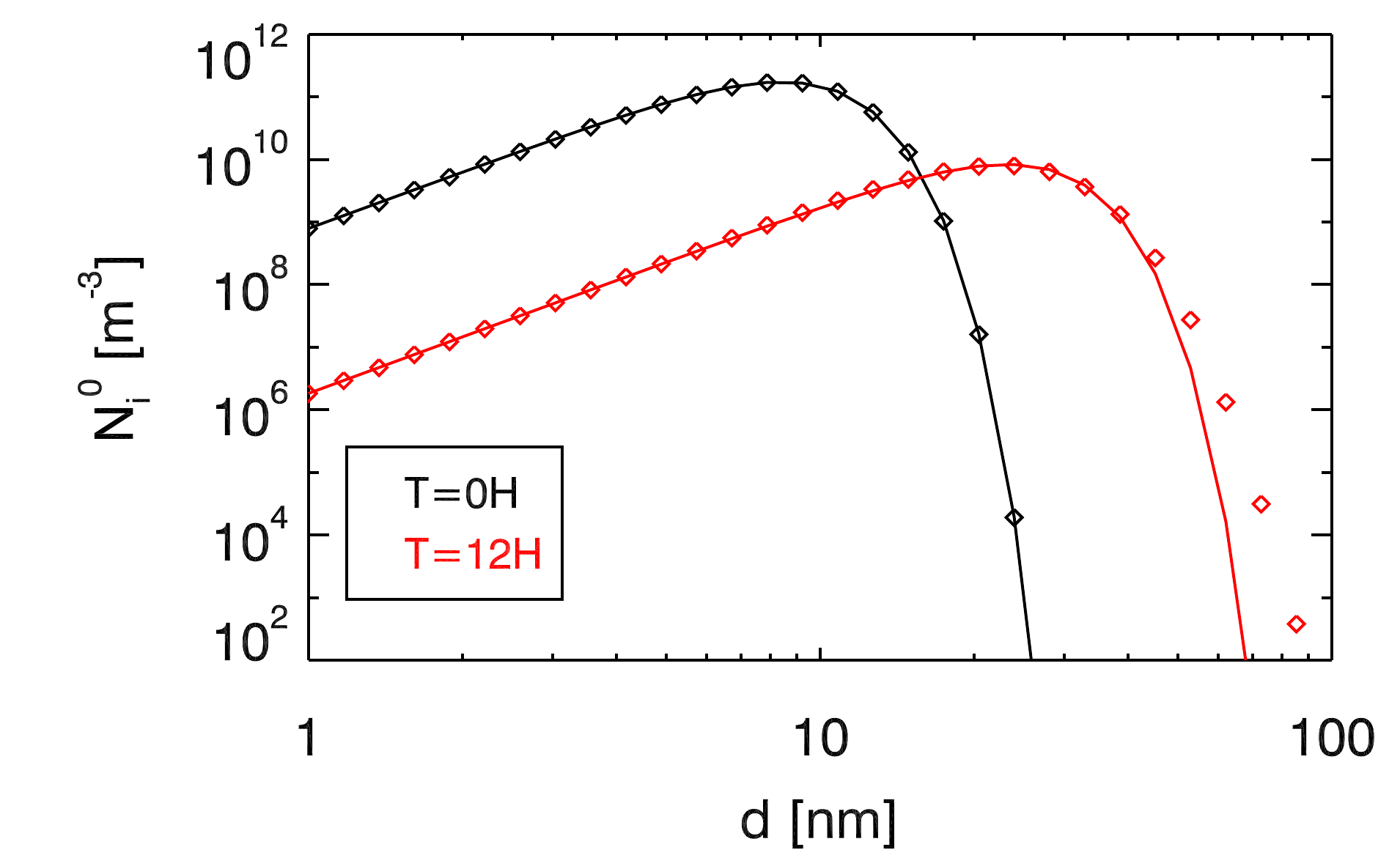}\hspace{-1.1cm}
        \includegraphics[width=7.6cm]{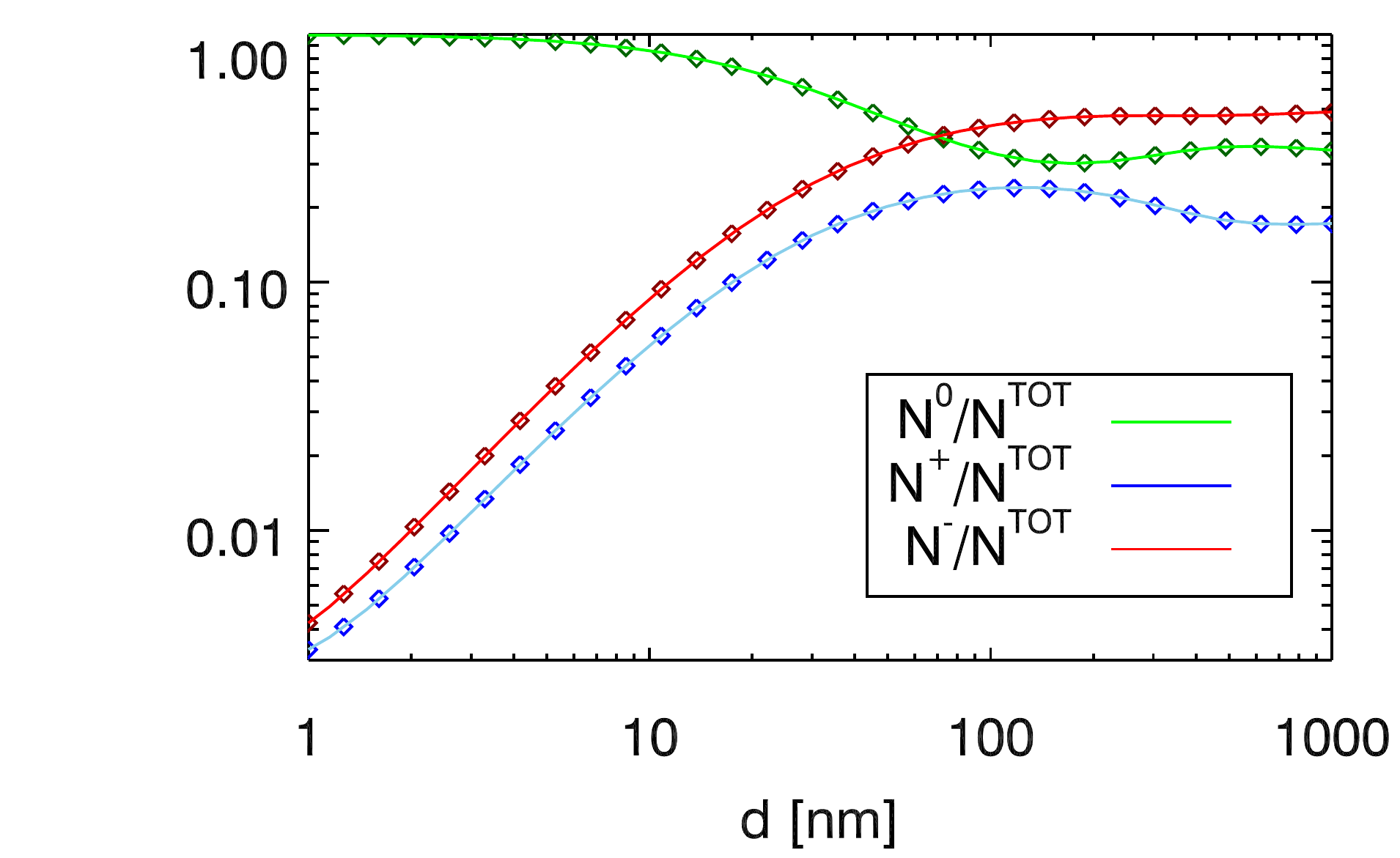} 
  \caption{Left: The particle spectrum initially (black) and after 12 hours of simulation for a distribution of the form Eq.\ \ref{eq:coag_analytic} subject to coagulation only, with a constant coagulation coefficient $K=10^{-15}\,\mathrm{m}^{2}\mathrm{s}^{-1}$. It can be noted that the analytic and model distributions begin to deviate for larger $D_p$ as time progresses. This is an effect of the logarithmic spacing of the size nodes, as can be seen also in Figure\ 15.2 of \cite{jacobson2005}. Right: Fraction of neutral, positively charged and negatively charged aerosols relative to the total number, as calculated by the model run in steady state (diamonds), and calculated analytically (lines). In this particular model run, the mass of the ions and are deliberately very different for demonstration purposes: The mass of the neutral monomer is $100\,$AMU, positive ion is $325\,$AMU and negative ion is $150\,$AMU. These masses are also reflected in the interaction coefficients, which is directly pronounced in the positive and negative aerosol fraction.}
   \label{fig:benchmark}
\end{figure}

We will now test the coagulation. The coagulation equation (Equation 13.72 of \citet{seinfeldpandis}), for a constant coagulation coefficient $\kappa_k^{qp}=K$ and initial continuous distribution 
\begin{equation}
    n(v,0) = \frac{N_0^2}{V_0}\exp\left(\frac{-vN_0}{V_0}\right)
\end{equation}
solves to
\begin{equation}\label{eq:coag_analytic}
n(v,t) = \frac{N_0^2}{V_0\left( 1+t/\tau_c\right)^2}    \exp{\left(       -\frac{vN_0}{V_0\left(1+t/\tau_c\right)}     \right)}.
\end{equation}
Note that now $n$ is a function of aerosol volume $v$. The characteristic time $\tau _c=2/(K N_0)$, $N_0$ is the number concentration of particles at the onset of the calculation or simulation, and $V_0$ is total volume concentration. The above analytic solution is compared with the numerical solution in Figure\ \ref{fig:benchmark} left panel. The black lines show the particle spectrum for $t=0$ as calculated by Eq.\ \ref{eq:coag_analytic}, and the black symbols show the simulated aerosol spectrum at the same instant. The red line and red symbols are the analytic and numerical solution evolved over $t=12$ hours. The behavior of the simulated spectra is seen to follow closely that of the analytic ones. It should be noted however, that as time progresses the largest particles coagulate beyond the upper size node, and as such the volume of simulated coagulation is not strictly conserved. This manifests itself as slight deviations from the analytic solutions towards the high end of the spectrum, and worsens as time progresses.

\subsection{Charge Distribution in Steady State}
A good benchmark for an aerosol model that handles charge is that the charged fractions of aerosols at a given volume node meet analytic expectations. The fraction of charged aerosols in a steady state scenario depends on the coefficients chosen for ion and aerosol interactions. These in turn are affected by a number of parameters, of which some are the ion masses. The ratio of the charged to neutral aerosol concentrations at a given volume node at steady state can be expressed as
\begin{eqnarray}
\frac{N_k^+}{N_k^0} = \frac{n^+\beta_k^{+0}}{n^-\beta_k^{-+}}, \\
\frac{N_k^-}{N_k^0} = \frac{n^-\beta_k^{-0}}{n^+\beta_k^{+-}},
\label{N+/N0}
\end{eqnarray}
which depends only on ion concentrations and condensation coefficients \citep{hoppel1985}. If $N_k^{\mathrm{tot}}=N_k^0+N_k^++N_k^-$ then the above equations solve to
\begin{eqnarray}
\frac{N^0_k}{N^{\mathrm{tot}}_k	} &=& \left( 1+\frac{n^- \beta_k^{-0}}{n^+ \beta_k^{+-}} + \frac{n^+\beta_k^{+0}}{n^-	 \beta_k^{-+}} \right)^{-1}, \nonumber  \\ 
\frac{N^+_k}{N^{\mathrm{tot}}_k	} &=& \left(1+\frac{n^-\beta_k^{-+}}{n^+ \beta_k^{+0}}\left(1 +\frac{n^-\beta_k^{-0}}{n^+ \beta_k^{+-}} \right)\right)^{-1},   \nonumber  \\
\frac{N^-_k}{N^{\mathrm{tot}}_k	} &=& \left(1+\frac{n^+\beta_k^{+-}}{n^- \beta_k^{-0}}\left(1 +\frac{n^+\beta_k^{+0}}{n^- \beta_k^{-+}} \right)\right)^{-1}.  \label{Eq:n0/ntot}
\end{eqnarray}
As an illustration, different condensation coefficients are chosen by setting the neutral monomer mass to $100\,$AMU, positive ion mass to $325\,$AMU and negative ion to $150\,$AMU (as opposed to later, where positive and negative ion masses are assumed equal). The model is then run into a steady state. The continuous lines of Figure\ \ref{fig:benchmark} in the lower right hand panel, show the analytic charge fractions given by Equation \ref{Eq:n0/ntot} as a function of aerosol diameter. The modeled charged aerosol fractions are shown as the diamond symbols, and demonstrate a good agreement between the analytical and modeled solutions. The model also reproduces the common observation that aerosols generally have a higher negative than positive charging state, due to differences in ion size and thus mobility \citep{Wiedensholer1988, Enghoff2017}, although the choice of masses in the present example may exaggerate this effect. Note that for larger aerosols (above $\sim$100$\,$nm) multiple charging starts to be relevant even though it is not included in the model.

\section{Case study: Simulated ion-induced condensation}
One of the unique features of the model presented in this work is the inclusion of the GDE terms describing ion induced condensation. It is of natural interest to quantify how this growth channel compares to neutral condensation, and this has indeed already been done in SESS17 taking a theoretical and experimental approach. Here it was shown that ion-condensation can be an important addition to the neutral condensation growth of aerosols under atmospheric conditions. It was found, that the addition of small concentrations of ions relative to the neutral condensable clusters heightened the probability of small nucleated aerosols surviving to cloud condensation nuclei (CCN) sizes $>50\,$nm, which will have an impact on cloud micro-physics in the terrestrial atmosphere. The strength of the theoretical ion-condensation description is that it shows how the ions contribution to growth rate is independent of aerosol size distribution. This description however neglects growth from coagulation, which may also be affected by the presence of ions. While the theoretical result of SESS17 is consistent with the experimental findings of the same paper, it is interesting to expand on aerosol growth from ion induced condensation in the context of neutral and charged coagulation. In order to do so, we first reconsider the theoretical description of aerosol growth in the context of charge. Our focus will lie on the growth rate
\begin{equation}
GR=\frac{d\,d}{dt}
\end{equation}
of aerosols at diameter $d$, and explore the conditions under which ion-induced condensation can be said to be important.

\subsection{Condensation growth rate}
In the case of purely neutral condensation, the growth rate can be written as 
\begin{equation}
GR_{\mathrm{cond}}^0 =  A_0 n^0 \beta^{00},
\end{equation}
where $A_0=(m^0/4\pi (d/2)^2\rho)$ is simply a constant related to the aerosol and monomer, such that $d$ is the aerosol diameter, $m^0$ the monomer mass and $\rho$ their density (SESS17). The $0$ superscript in the LHS indicates that only neutral condensation is considered. In the presence of ions, the contribution to the condensation is presented in SESS17 as
\begin{equation}
GR_{\mathrm{cond}}^{\pm} = A_0 n^0 \beta^{00}   \left(  1  +  \Gamma   \right),
\label{Eq:APPROX}
\end{equation}
where 
\begin{equation}
\Gamma =  4 \left(  \frac{n^\mathrm{\pm}}{n^0}   \right)   \left(   \frac{  \beta^{\pm 0}}{  \beta^{00}}  \right)   \left(  \frac{m^\mathrm{\pm}}{m^0} \right)  \left (  \frac{N^0(d)}{N^{\mathrm{tot}}(d)} \right). 
\label{Eq:gamma}
\end{equation}
$\Gamma$ is then the increase in growth rate caused by ion-condensation relative to the neutral growth rate for pure neutral condensation. Positive and negative ions are treated symmetrically, such that interaction coefficients for the two species are $\beta^{+0} = \beta^{-0} = \beta^{\pm 0}$, $m^+=m^-=m^{\mathrm{\pm}}$ and $n^+ = n^- = n^{\mathrm{\pm}}$.  The final factor $N^0/N^{\mathrm{tot}}$ is given in Equation \ref{Eq:n0/ntot} assuming charge equilibrium. Note that all $\beta$ are functions of aerosol diameter as shown in Figure \ref{fig:beta_II}. In Figure \ref{fig:K} on the left $\Gamma$ can be seen for a charge equilibrated distribution of aerosols with sulfuric acid monomers $n^0=10^6\,\mathrm{cm}^{-3}$, $m^0=100\,$AMU, $m^{\mathrm{\pm}}=225\,$AMU for a couple of ion-pair concentrations. As expected, $\Gamma$ scales with $n^{\pm}$.

\subsection{Coagulation growth rate}
While the result of the condensation above is universally applicable to any aerosol distribution, a similar expression is harder to achieve for the coagulation, where aerosols of all diameters may coagulate with each other. Here we shall consider the growth rate of a mono-disperse distribution for simplification, and then compare with simulations that are reasonably approximated as such. From Equation 9 of \citet{leppa2011}, the growth rate of a monodisperse concentration $N^T$ of completely neutral aerosols due to coagulation is 
\begin{equation}
GR_{\mathrm{coag}}^0 = \frac{d\,k^{00}N^T}{6}.
\end{equation}
Here $k^{00}(d)=\kappa^{00}(d,d)$ is the neutral-neutral coagulation coefficient of particles at equal diameters. $k^{00}$ is thus a function of diameter. Introducing ions, a fraction of the monodisperse aerosols will obtain a single charge. In this case, the growth rate from coagulation is
\begin{equation}\label{eq:leppa_charged_gr}
GR_{\mathrm{coag}}^{\pm} = \frac{d}{3N^T}\left[
\frac{1}{2}k^{00}(N^0)^2+
k^{0+}N^0N^{+}+
k^{0-}N^0N^{-}+
k^{+-}N^{+}N^-\right].
\end{equation}
As in the model, the terms with interactions between two positive aerosols and two negative aerosols have been neglected. Again all coagulation coefficients $k^{0+}$, $k^{0-}$ and $k^{+-}$ are functions of $d$, and now $N^T=N^0+N^++N^-$. If we assume that $k^{0\pm}=k^{0-}=k^{0+}$, and $N^\pm = N^+ = N^-$ then 
\begin{eqnarray}
GR_{\mathrm{coag}}^{\pm}&=&\frac{d}{3N^T}\left[
\frac{1}{2}k^{00}(N^0)^2+
2k^{0\pm}N^0N^{\pm}+
k^{+-}(N^{\pm})^2\right]\\
&=& 
\frac{d\,k^{00}N^T}{6}\left[\left(\frac{N^0}{N^T}\right)^2+
4\frac{k^{0\pm}}{k^{00}}\left(\frac{N^\pm}{N^T}\frac{N^0}{N^T}\right)+
2\frac{k^{+-}}{k^{00}}\left(\frac{N^{\pm}}{N^T}\right)^2\right]\\
&=& 
\frac{d\,k^{00}N^T}{6}\left(1+\Lambda\right),
\end{eqnarray}
where
\begin{equation}
    \left(1+\Lambda\right) = \left[\left(\frac{N^0}{N^T}\right)^2+
4\frac{k^{0\pm}}{k^{00}}\left(\frac{N^\pm}{N^T}\frac{N^0}{N^T}\right)+
2\frac{k^{+-}}{k^{00}}\left(\frac{N^{\pm}}{N^T}\right)^2\right].
\end{equation}
The relative concentrations of the $1+\Lambda$ expression can be calculated based on Equation \ref{Eq:n0/ntot}
assuming the aerosols to be in equilibrium with respect to their charge. Then $\Lambda$ is independent of $N^T$ and $n^{\pm}$. Thus the GR from coagulation has a unit-less and diameter dependant adjustment $\Lambda(d)$ that is given by the coagulation coefficients and equilibrium charge fractions. $\Lambda$ can be seen plotted in Figure \ref{fig:K} on the left. For $d>100\,$nm, $\Lambda$ actually becomes slightly negative, since the assumption of like-charged coagulation is neglected. Including the like-charged terms of Equation \ref{eq:leppa_charged_gr} would produce an increase $\widetilde{\Lambda}$ as seen in Figure \ref{fig:K}, and for more advanced treatments multiple charge aerosol species could be included in the term. The fraction of multiply charged aerosols is however low for sizes below 100$\,$nm, going from $\approx$1\% to $\approx$5\% between $d=40\,$nm and $d=100\,$nm \cite{hoppel1986}. As multi-charge aerosols are not handled by the model of the present work, we exclude it for consistency and direct comparison and use only $\Lambda$. 

\subsection{Growth rate increase from ions}
With expressions for the neutral and charged growth rate from condensation and coagulation in hand, we can consider two cases: 1) The growth rate due to both coagulation and condensation for a neutral monodisperse concentration $N^T$ of aerosols, and 2) The growth rate of the same concentration $N^T$ of aerosols however with the presence of ions, condensing and charging a fraction of the aerosols. Taking the ratio of the two growth rates we obtain a measure of the increase in growth rate due to the presence of charging ions on aerosols at diameter $d$ and concentration $N^T$:

\begin{figure}
   \centering
        \includegraphics[width=14cm]{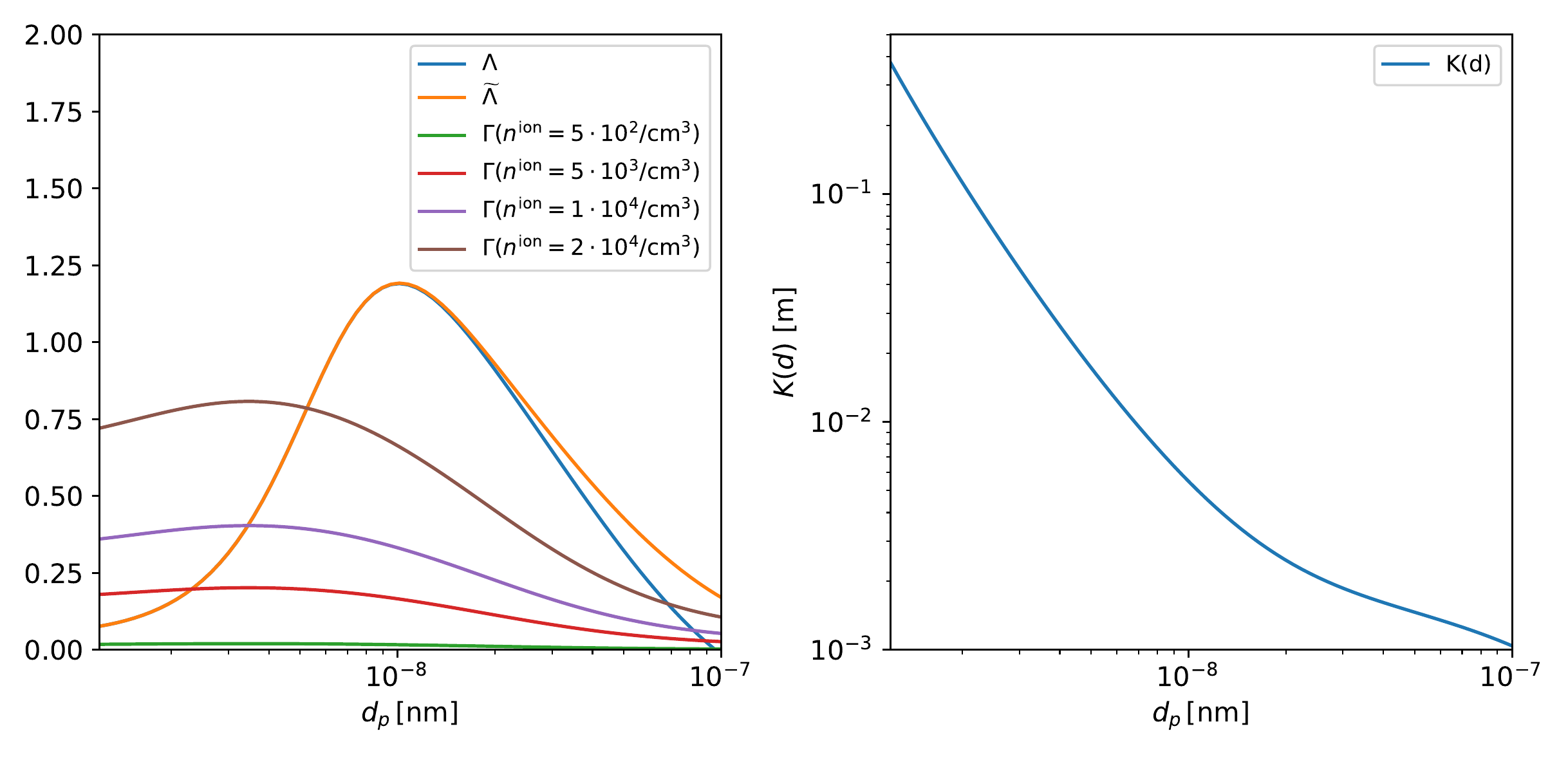}
  \caption{\textit{Left:} The $\Lambda$ and $\widetilde{\Lambda}$ profiles associated with charged coagulation, and $\Gamma$ profiles for a number of $n^{\pm}$ values associated with ion induced condensation \textit{Right}: $K(d)$ from Equation \ref{eq:GR}.}
   \label{fig:K}
\end{figure}

\begin{eqnarray}\label{eq:GR}
    \Omega &=& \frac{GR_{\mathrm{cond}}^{\pm} + GR_{\mathrm{coag}}^{\pm}}
    {GR_{\mathrm{cond}}^0 + GR_{\mathrm{coag}}^0} \\ \nonumber
    &=&
    \frac{K(d)\frac{n^0}{N^T}(1+\Gamma)+1+\Lambda}{K(d)\frac{n^0}{N^T}+1},
\end{eqnarray}
where $K(d)=\frac{6A_0(d)}{d}\frac{\beta^{00}(d)}{k^{00}(d)}$. $K(d)$ can be seen in Figure \ref{fig:K}, and ranges from $\approx 10^{-2}\,$m at $d=5\,$nm to $\approx 10^{-3}\,$m at $d=100\,$nm.

To explore the regimes in which coagulation or condensation dominates the GR increase due to ions, we explore the limits of Equation \ref{eq:GR}. If we assume $N^T<<K(d)n^0$ condensation is the dominant growth mechanism. $\Omega$ reduces to
\begin{equation}\label{eq:condonly}
    \Omega = 1+\Gamma.
\end{equation}
On the other hand, if  $N^T>>K(d)n^0$ then the coagulation is the dominant growth mechanism, and the expression reduces to 
\begin{equation}\label{eq:coagonly}
    \Omega = 1+\Lambda.
\end{equation}
In the next subsection we simulate these two cases for appropriate atmospheric values, as well as cases in between, and compare with the equations above.

\subsection{Growth rates in simulation}
In order to calculate growth rates from the model, an initial number $N^T$ of aerosols at $d=1\,$nm were grown in a loss-free environment with fixed sulfuric acid concentration $n^0$ and fixed ion concentration $n^{\mathrm{\pm}}$, using parameters as found in Table \ref{tab:parameters_SESS}. As our goal is to compare model and theoretical results, two consideration must be made: 1) We focus on the simulation mean diameter rather than GR for the entire distribution, and 2) when considering coagulation we must take into consideration the fact that the number of aerosols $N^T$ changes dynamically in each simulation depending on ambient conditions.

Since the model suffers from numerical diffusion as well as a broadening of the originally narrow distribution due to coagulation itself, the simulation quickly departs from the mono-disperse initial conditions assumed above, so some extra care is needed to compare to the expressions in the previous subsection. At each point in time through each simulation we can calculate the mean aerosol volume in the simulation
\begin{equation}
    \left<v\right> = \frac{\Sigma_k N_k^Tv_k }{\Sigma N_k^T},
\end{equation}
from which the mean diameter $\left<d\right>$ is obtained. We can then consider the growth rate as 
\begin{equation}
    GR = \frac{d\left<d\right>}{dt}.
\end{equation}
The equivalent number of aerosols at this diameter is simply $N^T$:
\begin{equation}
    N^T = \frac{\Sigma_k N_k^Tv_k }{\left<v\right>}=\Sigma_k N_k^T.
\end{equation}
Assuming $d=\left<d\right>$ enables us to compare the simulated growth rate of the mean diameter to that of the analytical mono-disperse distribution above. Then, we can produce simulated $\Omega$ by comparing the GR in simulations with similar $N^T$ at similar $\left<d\right>$ with and without ions present. 

\begin{figure}
   \centering
        \includegraphics[width=13cm]{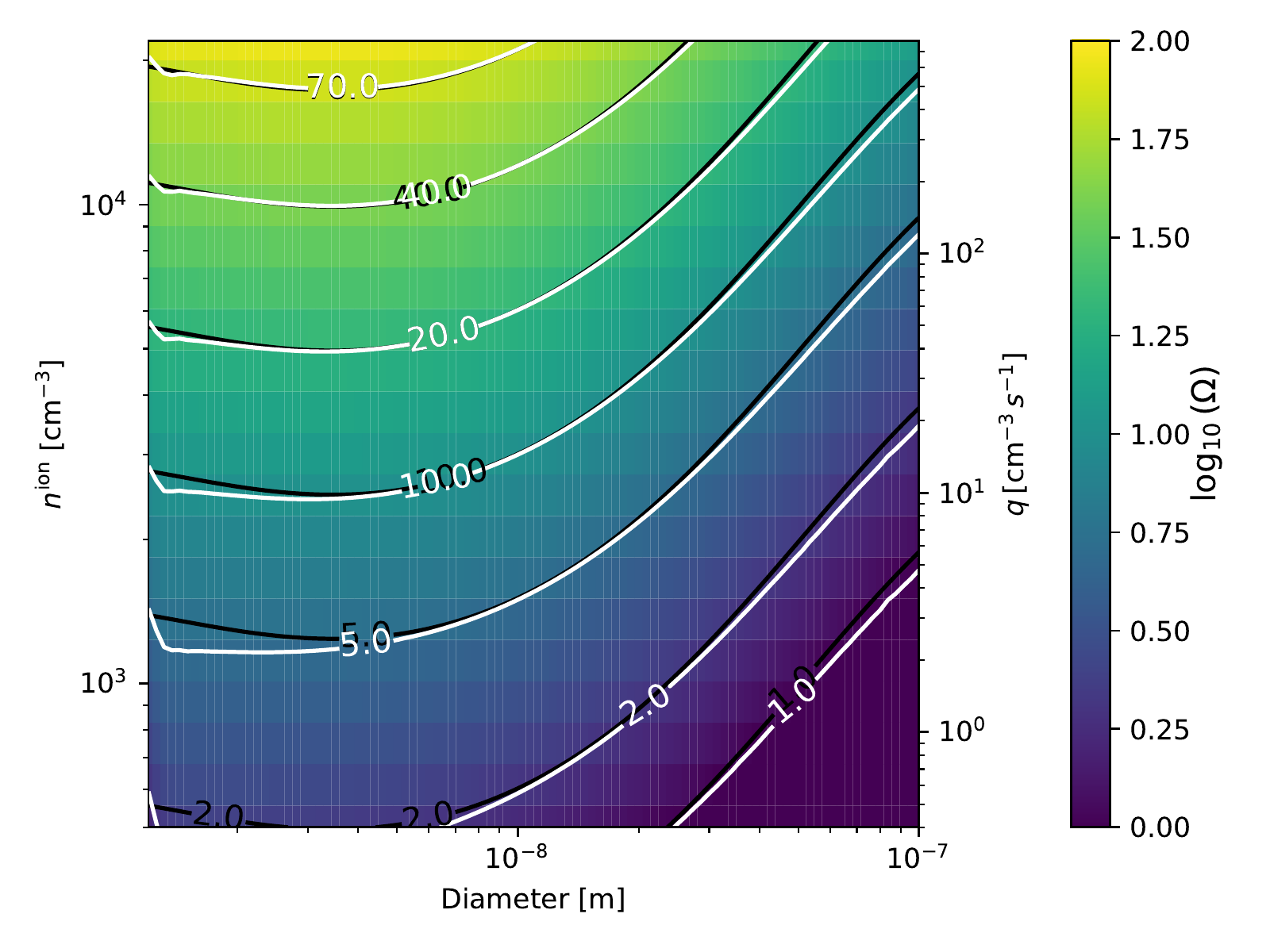}
        \includegraphics[width=13cm]{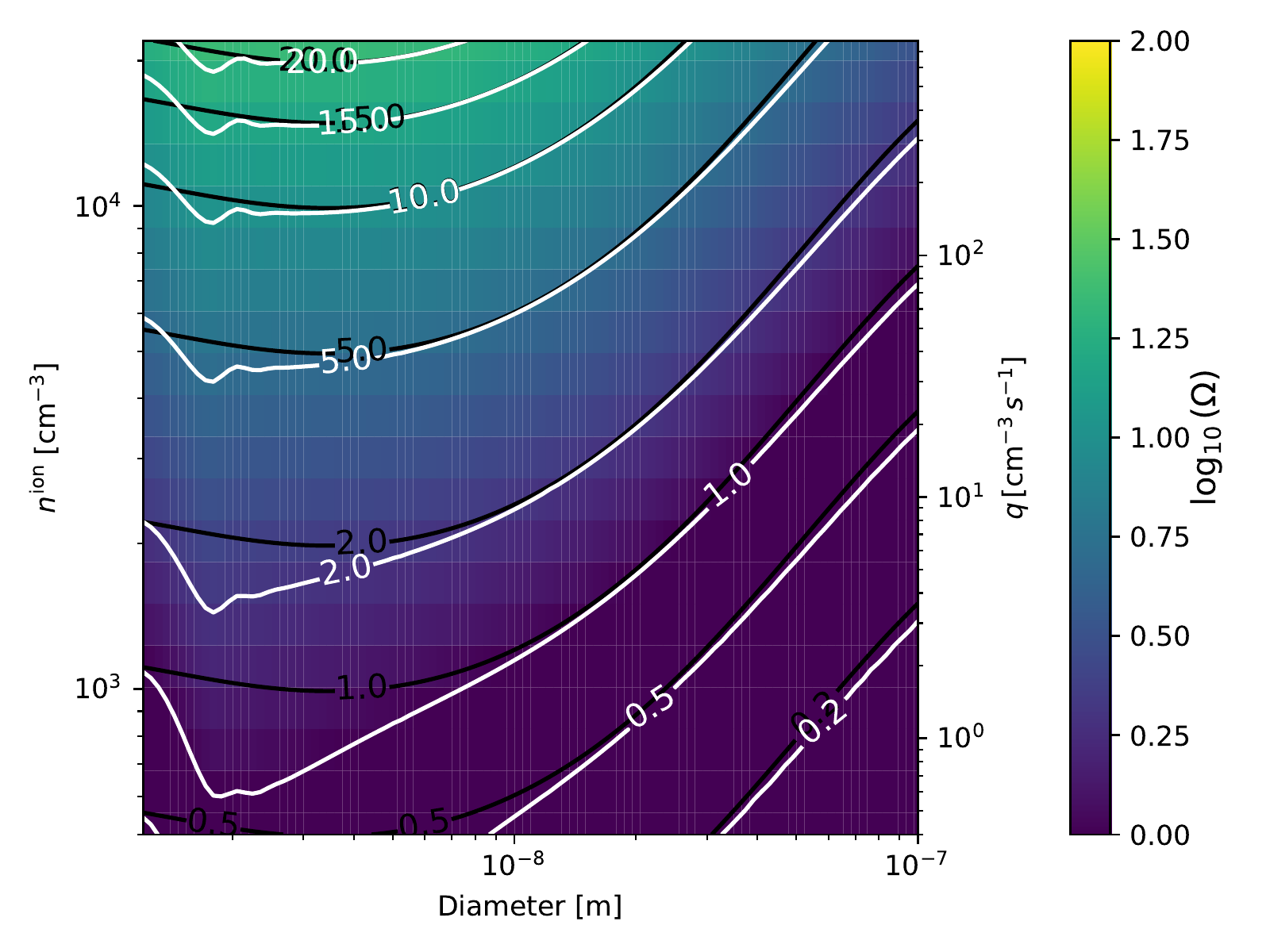}
  \caption{Charge enhancements of the growth rate $\Omega$ from condensation only. The colors represent the simulated growth rates, and the white contours denote the same, in percent. The black contours show the theoretical contours of equation \ref{eq:GR}. The $y$-axis on the left hand side of the panels shows the fixed ion-pair concentration, and the right hand side shows the equivalent ion-pair production rate as calculated by $q=1.6\times10^{-6}(n^{\pm})^2$ \textit{Top}: $n^0=1\times10^{-6}\,\mathrm{cm}^{-3}$, equivalent to Figure 1 of SESS17. \textit{Bottom}: $n^0=4\times10^{-6}\,\mathrm{cm}^{-3}$}
  \label{fig:cond_only_simulation}
\end{figure}

First, we focus on the condensation only-case of Equation \ref{eq:condonly}. Given a fixed ion-pair concentration $n^{\pm}$, an initially monodisperse distribution of $N^T=1000\,\mathrm{cm}^{-3}$ was grown towards higher diameters. Growth rates of the mean diameter in simulations of 20 different levels of $n^{\pm}$  was compared to the growth rate at the equivalent diameter of the neutral ($n^{\pm}=0$) case for two levels of sulfuric acid $n^0=1\times 10^6\mathrm{cm}^{-3}$ and $n^0=4\times 10^6\mathrm{cm}^{-3}$. Both of these can be seen in Figure \ref{fig:cond_only_simulation}. The upper panel is equivalent to Figure 1 of SESS17. There is an excellent correspondence between the expression for $\Omega=1+\Gamma$ and the simulated GR enhancement in a coagulation free environment. The effect scales with $n^{\pm}/n^0$, and by quadrupling $n^0$ from the top to the bottom panel in Figure \ref{fig:cond_only_simulation}, the effect is seen to be suppressed exactly with this factor in response. This is in effect the ion condensation effect.

\begin{figure}
   \centering
        \includegraphics[width=11cm]{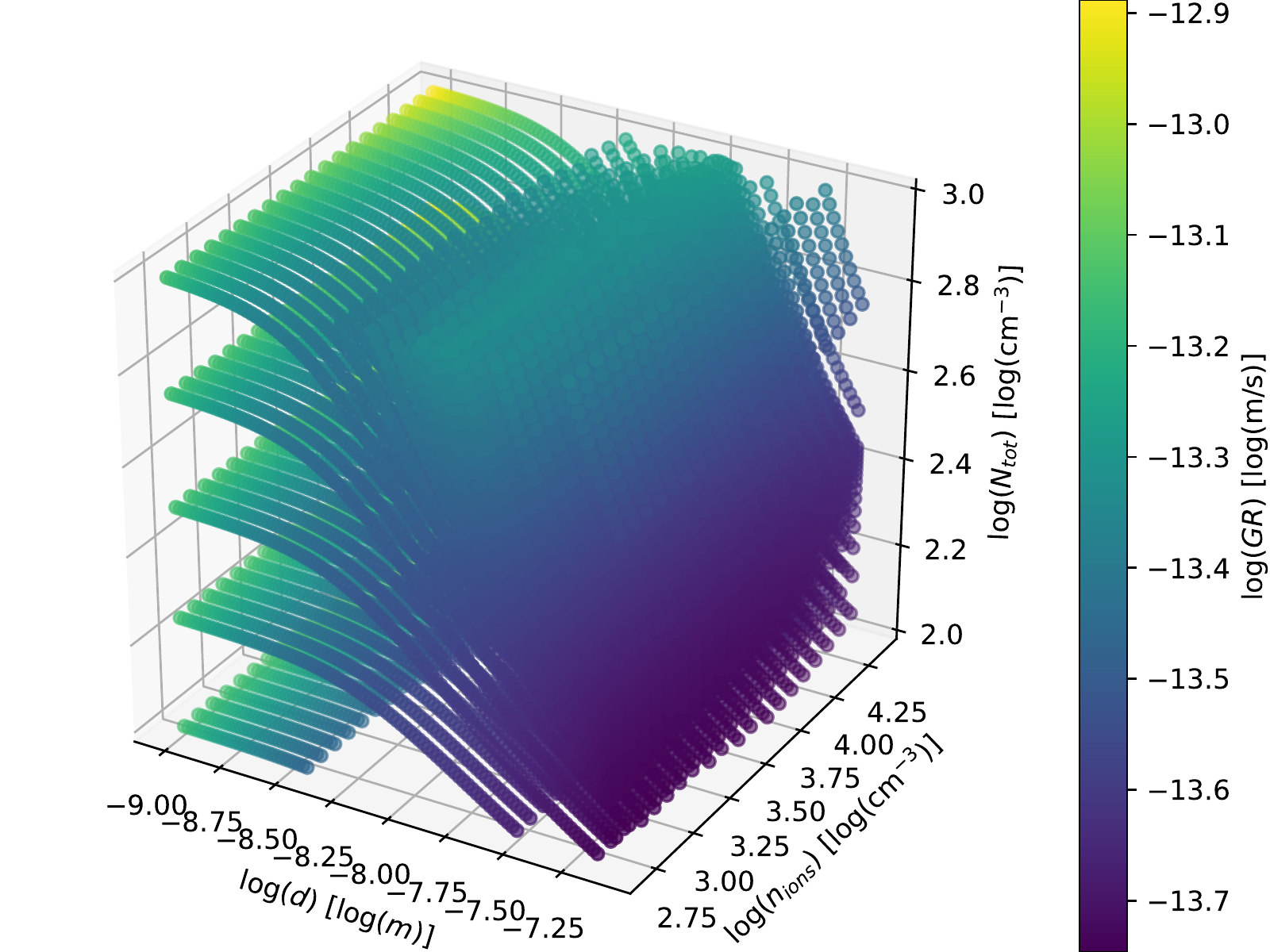}
  \caption{Subset of mean diameter growth rates in simulations at constant $n^{ion}$, including condensation and coagulation growth terms. As $N^T$ decreases in a given simulation due to coagulation.}
   \label{fig:SIMS}
\end{figure}

Now, including also coagulation Equation \ref{eq:GR} shows how $\Omega$ a function of $d$, $n^{\pm}$, $n^0$ and $N^T$. Within the course of a single simulation, as the aerosols coagulate $N^T$ decreases, which in turn reduces the effect of coagulation. It is convenient to be able to keep the aerosol concentration $N^T$ fixed while comparing growth rates at varying $d$, $n^{\pm}$ and $n^0$. Therefore, given an ion-pair concentration $n^{\pm}$, 20 simulations of initial $N^T$ aerosols between $10^2\,$cm$^{-3}$, and $10^8\,$cm$^{-3}$ were computed. Then, even though $N^T$ decreases with time within a single simulation, the growth rate of the mean diameter at concentration $N^T$ could be interpolated between simulations. We assume a value of $n^0=10^6\,\mathrm{cm}^{-3}$, and for around 20 values of $n^{\pm}$ between 0 and 22005$\,$cm$^{-3}$ we allow a concentration of aerosols at 1.1$\,$nm to coagulate, condensate, and interact with ions. For each simulation, the growth rate $GR(\left<d\right>)$ and total aerosol concentration $N^T(\left<d\right>)$ of the mean diameter $\left<d\right>$ is obtained. A subset of the simulation growth rates are seen in Figure \ref{fig:SIMS}. The trajectory of $\left< d\right>$ in this space for a single simulation can be seen as beginning at the lowest value of $\left< d\right>$, for an initial $N^T$, and growing towards higher values as time progresses while $N^T$ decreases. 

\begin{figure}
   \centering
        \includegraphics[width=13cm]{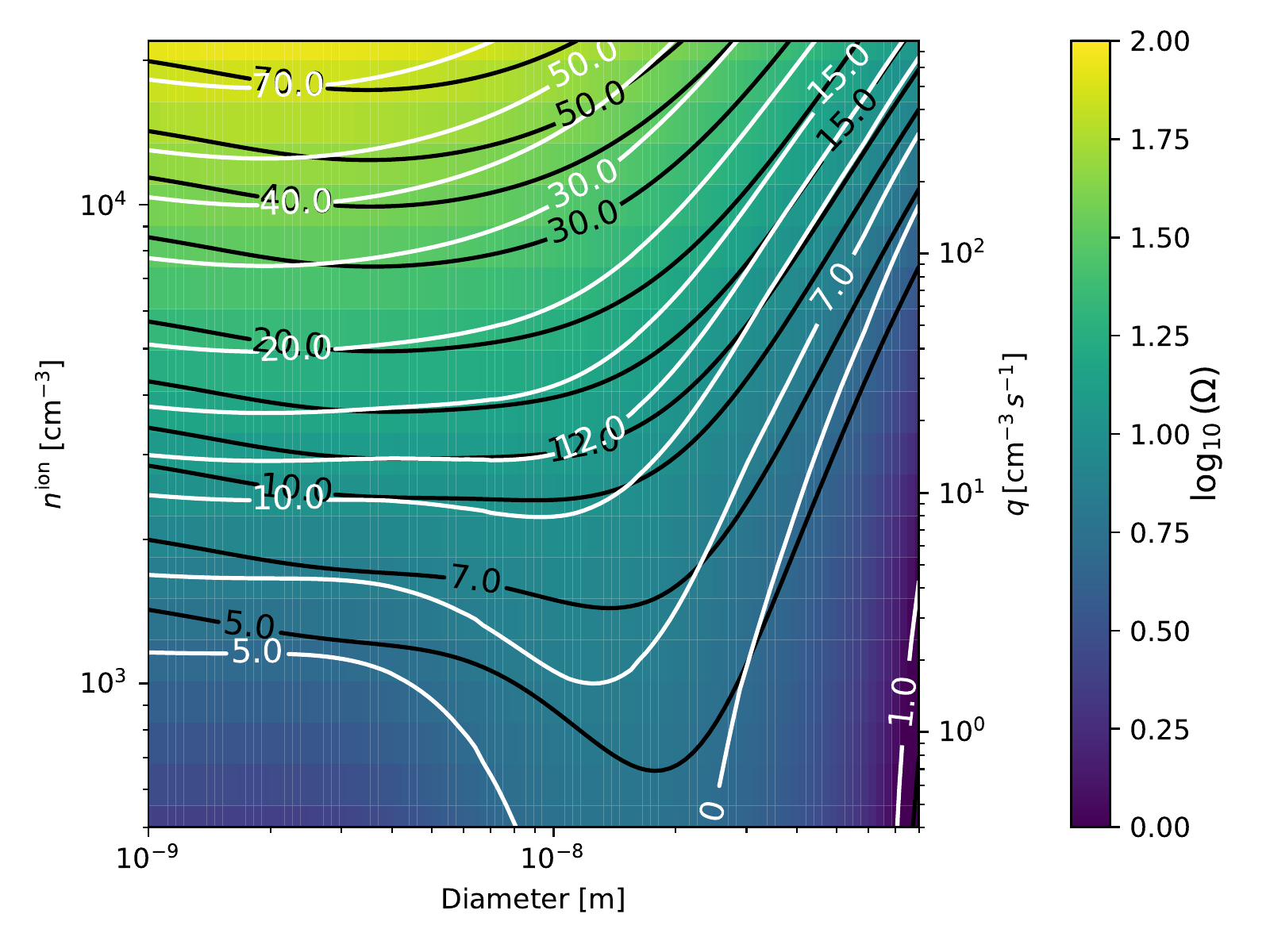}
        \includegraphics[width=13cm]{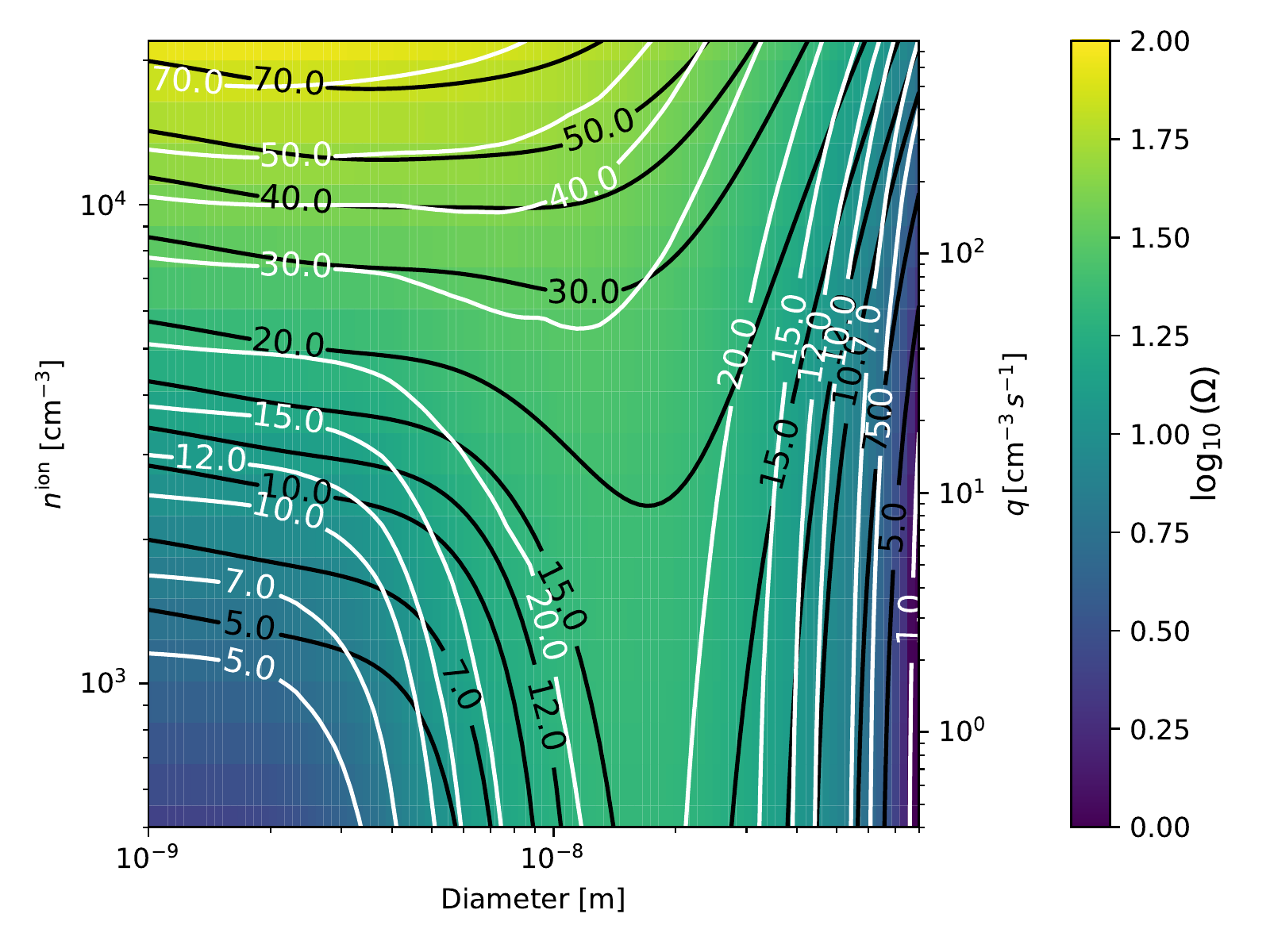}
  \caption{Charge enhancements of the growth rate $\Omega$ from condensation and coagulation. The colors represent the simulated growth rates, and the white contours denote the same, in percent. The black contours show the theoretical contours of Equation \ref{eq:GR}. Both panels are calculated from a horizontal slice through the data of Figure \ref{fig:SIMS}. The $y$-axis on the left hand side of the panels shows the fixed ion-pair concentration, and the right hand side shows the equivalent ion-pair production rate as calculated by $q=1.6\times10^{-6}(n^{\pm})^2$ \textit{Top}: $N^T=100\,\mathrm{cm}^{-3}$. \textit{Bottom}: $N^T=500\,\mathrm{cm}^{-3}$}
  \label{fig:cond_coag_simulation}
\end{figure}

Interpolating in this space for two constant values of $N^T$ yields the two panels of Figure \ref{fig:cond_coag_simulation}. Here, the added effects of ions on the coagulation and condensation i.e. $\Lambda$ and $\Gamma$ are comparable in size. As can be seen when comparing the pure condensation in Figure \ref{fig:cond_only_simulation} and pure coagulation contained in $\Lambda$ of Figure \ref{fig:K} with Figure \ref{fig:cond_coag_simulation}, we can identify the two components that contribute to the charged growth rate enhancement. Ion induced condensation enhances growth rates for increasing $n^{\pm}$ and small diameters i.e. the bulge in the upper left corners. Added growth from charged coagulation via $\Lambda$ manifests itself as an increase with diameter, followed by a decreases back to uncharged growth as coagulation coefficients for single-charge aerosols and neutral aerosols tend to be equal for very large diameters. This corresponds to the vertical band across the two figures. The coagulation effect is independent of $n^{\pm}$, as expected when aerosols are in a charged fraction equilibrium. In the top panel, a constant value of $N^T=100\,\mathrm{cm}^{-3}$ is depicted. In the bottom panel, $N^T=500\,\mathrm{cm}^{-3}$, and as expected, the contribution to $\Omega$ from coagulation is stronger while the condensation of ions remains the same. There is a good, but not 1-to-1 correspondence between the simulated and theoretical contours in the two figures, however the principal behaviour of theory and simulation is the same.

\begin{figure}
   \centering
        \includegraphics[width=13cm]{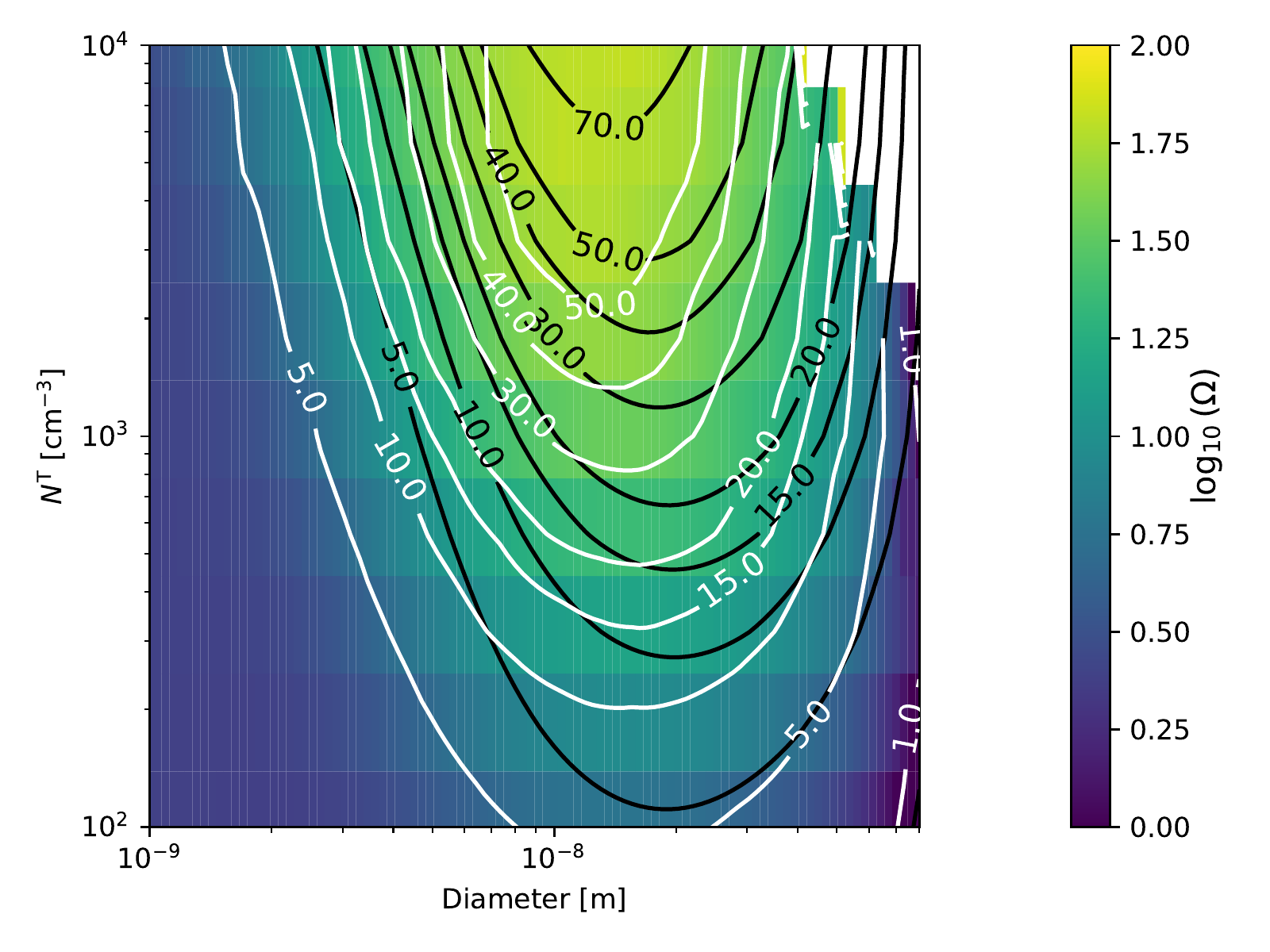}
        \includegraphics[width=13cm]{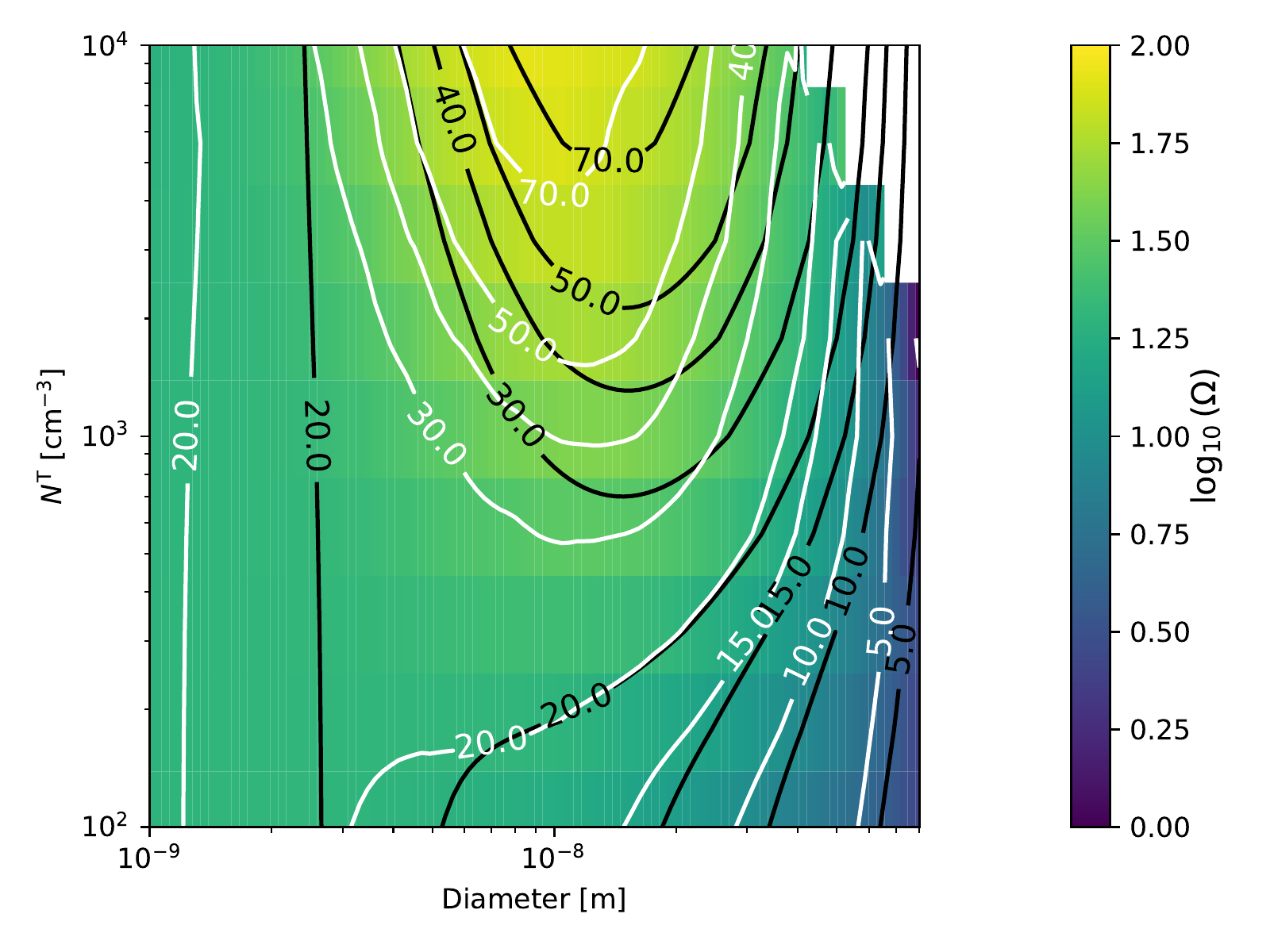}
  \caption{Charge enhancements of the growth rate $\Omega$ from condensation and coagulation, with a constant $n^{ion}$. The colors represent the simulated growth rates, and the white contours denote the same, in percent. The black contours show the theoretical contours of Equation \ref{eq:GR}. Both panels are calculated from a vertical slice through the data of Figure \ref{fig:SIMS}. The $y$-axis on the left hand side of the panels shows the fixed ion-pair concentration, and the right hand side shows the equivalent ion-pair production rate as calculated by $q=1.6\times10^{-6}(n^{\pm})^2$ \textit{Top}: $n^{\pm}=500\,\mathrm{cm}^{-3}$. \textit{Bottom}: $n^{\pm}=5000\,\mathrm{cm}^{-3}$}
  \label{fig:cond_coag_simulation_constant_ions}
\end{figure}

Holding instead $n^{\pm}$ as constant, and using vertical slices along the $d$ direction of Figure \ref{fig:SIMS} allows for the calculation of the data in Figure \ref{fig:cond_coag_simulation_constant_ions}. In comparison to Figure \ref{fig:cond_coag_simulation} $N^T$ is now varied instead of $n^{\pm}$. As $N^T$ increases, coagulation and therefore $\Lambda$ becomes progressively more dominant for $\Omega$. Increasing $n^{\pm}$ from the top panel to the bottom panel makes the $\Gamma$ ion condensation term contribute more to $\Omega$ for the lower diameter aerosols, on par with the theoretical $\Gamma$.

 \begin{table}
 \caption{Parameters for the model to reproduce the results of the SESS17 paper.}
 \centering
 \begin{tabular}{lll}
 \hline
Parameter & value & Description \\
 \hline
 $m^0$ & 100$\,$AMU & Sulfuric Acid \\
 $m^{+}$ & 225$\,$AMU & As found in the SESS17 paper\\ 
 $m^{-}$ & 225$\,$AMU & As found in the SESS17 paper\\
 $\rho^0$ & 1200$\,$kg$\,$m$^{-3}$ & Sulfuric Acid \\
 $\rho^+$ & 1200$\,$kg$\,$m$^{-3}$ & Sulfuric Acid \\
 $\rho^-$ & 1200$\,$kg$\,$m$^{-3}$ & Sulfuric Acid \\
 No nucleation terms & &\\
 No loss terms & & \\
 \hline
 \end{tabular}
\label{tab:parameters_SESS}
 \end{table}
 
It should be noted that in general, ion mass, concentrations and interaction coefficients can be set and calculated individually, in the code, and need not to be restricted to the symmetric case tested here.

\color{black}
\section{Discussion and Conclusion}
In the present work a numeric model for simulating aerosol growth in the presence of charge has been developed and presented. The code consists of a 0-dimensional time integration model, where the interactions between ions and (charged) aerosols are taken into account. In particular ions are treated as massive condensing particles contributing to the growth rate in an implementation similar to condensation of neutral clusters here modeled as (H$_2$SO$_4$), while accounting for their inherent transfer of charge. The model is then used to investigate the newly demonstrated effect of ion-induced condensation from \citet{Svensmark2017} and a good agreement between model output and the theoretical and experimental SESS17 results is demonstrated employing condensation-only assumptions. Once coagulation is taken into account, we explore the conditions under which ion condensation is relevant, and show that while coagulation adds to the growth rate of aerosols when ions are present, it is not dependent on ion concentration, and in this sense the ion-condensation contribution to growth rate can be said to function independent of coagulation. Naturally in both experimental and real atmosphere scenarios there will be growth due to coagulation when the aerosol distribution is not well approximated by monodisperse conditions, there will be loss mechanisms and temporally varying concentrations of the ion-pairs and neutral clusters from which aerosols grow in size. Furthermore the positive and negative ions may in general be of different masses. These are parameters and dynamics which may be handled by the model in its current implementation.

The concept of ion-induced condensation is still new, and further investigation into its dynamics and implications are needed. The present model may prove a useful tool for describing the mechanism under a range on conditions and parameters relevant for the atmosphere, including some that would otherwise be unobtainable in the laboratory. While we have done a step towards describing it in the context of aerosol coagulation, more detailed work remains for future studies.

The model could also be expanded to include more or different cluster-species. This could be relevant for investigating, for instance, how these species affect the growth of aerosols such as the role of organics in early aerosol growth \citep{Trostl2016}. Implementing evaporation of small clusters would allow for the model to be used in nucleation studies as well for systems relevant to Earth \citep{Zhang2004} or more exotic systems like TiO$_2$ relevant to brown dwarfs or exoplanets \citep{Lee2015}. Further more, the model may provide insight into the fast dust grain growth observed in supernovae remnants \citep{gall2014, bevan2017} and in high redshift galaxies \citep{watson2015,michalowski2015}.

%
%
%
\newpage

\appendix

\section{Interaction parameters between particles}
\label{app:coeff}
Here the parameters of the model are defined. We note that the calculated enhancement factors and potentials are system specific and that the calculations presented here should be considered as a first approximation.

\subsection{Brownian coagulation kernel}
The interaction coefficient between two aerosols in the Brownian regime is given by \citet{Jacobson2004} as
\begin{equation}
{\beta^B}_{i,j} =  \frac{4 \pi (a_{i}+a_{j})(D_{m,i}+D_{m,j})}{\frac{a_{i}+a_{j}}{a_{i}+a_{j}+(\delta_{m,i} + \delta_{m,j})^{1/2}}+\frac{4(D_{m,i}+D_{m,j})}{\sqrt{\overline{v}^2_i+\overline{v}^2_j}(a_{i}+a_{j})}} V_{i,j},
\end{equation}
where $a_i$ and $a_j$ are the radii of the two interacting particles. $\overline{v}_i  = \sqrt{8  k_B T/(\pi M_i)}$ is the average speed of the aerosol, with $M_i$ being the mass of the aerosol. $D_{m,i}$ is the diffusion constant of particle of radius $a_i$
\begin{equation}
D_{m,i} = \frac{k_B T}{6 \pi a_i \eta}  \left[1 + \mathrm{Kn}_i  \left( A_m+B_m \exp(-C_m/\mathrm{Kn}_i)\right) \right],
\end{equation}
where the Knudsen number is given by $\mathrm{Kn}_i =  \lambda / a_i $ and where $\lambda$ is the mean free path of an air molecule.  $\eta$ is the dynamic viscosity of air, and  
 \begin{equation}
 \delta_i = \frac{(2 a_i+\lambda_i)^3 - (4 a_i^2 + \lambda_i^2)^{3/2} }{ 6 a_i \lambda_i} - 2  a_i.
\end{equation}
The three constants in the diffusion term are quoted by \cite{kasten1968} as
\begin{eqnarray}
A_m &= &  1.249,  \nonumber \\ 
B_m &= & 0.42, \nonumber \\ 
C_m &= &  0.87. \nonumber
\end{eqnarray}
Finally the $V_{i,j}$ is an enhancement factor that can take interactions between the aerosols into account. This term will be defined in the next section. 

\subsection{Calculation of enhancement factors for coagulation of charged ions}
In the case of charged aerosols, the Brownian coagulation coefficients should be corrected since their charge as well as their subsequent induction of additional charge results in a modified potential. For any potential, the correction factor for aerosols in the continuum regime ($Kn \rightarrow 0$) of radii $a_i$ and $a_j$ is \citep{Jacobson2004},
\begin{equation}
W^c_{i,j} = \frac{1}{(a_i+a_j)\int_{a_i+a_j}^{\infty} D_\infty/D_{i,j}  \exp\left[\frac{E_{i,j}(r)}{k_BT}\right]r^{-2}dr },
\end{equation}
where $E_{i,j}(r)$ is the interaction potential between the two aerosols at separation $r$, $k_B$ is the Boltzmann constant and $T$ is the temperature. Viscous forces are taken into account by the following expression
\begin{equation}
D_\infty/D_{i,j} = 1+ \frac{2.6 ~ a_i a_j}{(a_i+a_j)^2} \sqrt   \frac{a_i a_j}{ (a_i+a_j) (r-a_i-a_j)} +\frac{a_i a_j}{ (a_i+a_j) (r-a_i-a_j)}. 
\end{equation}

In the kinetic regime ($Kn \rightarrow \infty$) when the interaction between the two aerosols is attractive and has a singularity at contact, the enhancement factor is \citep{Marlow1982}
\begin{equation}
W^k_{i,j} = \frac{-1}{2(a_i+a_j) k_B T} \int_{a_i+a_j}^{\infty} \left ( \frac{dE_{i,j}}{dr}   +  r \frac{d^2E_{i,j}}{dr^2} \right)  \exp \left [  \frac{-1}{k_B T}  \left (   \frac{r}{2}  \frac{dE_{i,j}}{dr}  + E_{i,j}   \right )  \right ]   r^2   dr.
\end{equation} 
\subsection{The potentials}
In the present work the aerosols can be charged with a single charge, either positive or negative. Since the aerosols are made of dielectric medium the charge interaction will involve mirror charges, and the interaction potential will be slightly more involved. The potential representation used here is based on \citet{Bichoutskaia2010} for two spheres of radii $a_i$ and $a_j$. For the two spheres with charge $Q_i$ and $Q_j$, permittivity $\epsilon _i$ and $\epsilon _j$ and at a central distance $r$, 
\begin{eqnarray}
E_{EM,i,j}(r) &= &K\frac{Q_iQ_j}{r}-Q_i\sum_{m=1}^{\infty}\sum_{l=0}^{\infty} 
                  \frac{(k_j-1)m}{(k_j+1)m+1}\frac{(l+m)!}{l!m!} \nonumber \\
                  &\quad\quad\times&\frac{a_j^{2m+1}}{r^{2m+l+2}}A_{1,l}
                  -\frac{1}{K}\sum_{l=1}^{\infty}\frac{(k_i+1)l+1}{(k_i-1)l}\frac{A_{1,l}A_{1,l}}{a_i^{2l+1}},
\end{eqnarray}
where $K= 1/4 \pi \epsilon_0 \approx  9\cdot10^9 $ Vm/C, $k_i=\epsilon _i/\epsilon _0$ and $k_j=\epsilon _j/\epsilon _0$ and $A_{1,l}$ are the coefficients found by solving the following implicit system of linear equations:
\begin{eqnarray}
A_{1,l} &= & a_iV_i\delta_{l,0}-
                \frac{(k_i-1)l}{(k_i+1)l+1}\frac{a_i^{2l+1}}{r^{l+1}}a_jV_j
               +\frac{(k_i-1)l}{(k_i+1)l+1}\nonumber\\
               &\quad\quad\times & \sum_{l_2=0}^{\infty}\sum_{l_3=0}^{\infty}
               \frac{(k_j-1)l_2}{(k_j+1)l_2+1}\frac{(l+l_2)!}{l!l_2!}\frac{(l_2+l_3)!}{l_2!l_3!}\nonumber\\
              & \quad\quad\times & \frac{a_i^{2l+1}a_j^{2l_2+1}}{r^{l+2l_2+l_3+2}}A_{1,l_3}.
\end{eqnarray}
Here $a_i V_i = KQ_i$, $a_j V_j = KQ_j$, and $\delta_{l,0}$ is the Kronecker delta function.

In addition to the electrostatic potential, it is assumed that there are short range Van der Waals forces present. This potential will ensure that there is a singular potential at contact and make the expression for the kinetic enhancement factor valid \citep{Marlow1982}. The Van der Waals potential is given by
\begin{equation}
E_{\mathrm{vdW},i,j}(r) =  -\frac{A}{6} \left[ \frac{2a_ia_j}{ r^2-(a_i+a_j)^2} + \frac{2a_ia_j}{r^2-(a_i-a_j)^2} + 
\log \left( \frac{r^2-(a_i+a_j)^2}{r^2-(a_i-a_j)^2} \right) \right].
\end{equation}
Here $A$ is the Hamaker constant measured to be  $A  = (6.4 \pm 2.6)  10^{-20}\,$J for H2SO4-H2O aerosol interactions at 300$\,$K \citep{CHAN2001}.

\subsection{Connecting the continuum regime with the kinetic regime}
Calculation of  $W^c_{i,j}$ and $W^k_{i,j}$ must be done numerically and it is therefore convenient to make the following variable transformation  
\begin{equation}
x = \frac{a_i+a_j}{r}.
\end{equation}
The interval of integration is changed from $[a_i+a_j; \infty[$ to $[0;1]$, so that the enhancement factor in the continuum limit becomes
\begin{equation}
W^c_{i,j} = \left ({\int_{0}^{1} D_\infty/D_{i,j} \exp\left[\frac{E_{i,j}((a_i+a_j)/x)}{k_BT}\right] dx } \right)^{-1},
\end{equation}
while the enhancement factor in the kinetic limit becomes
\begin{equation}
W^k_{i,j} = \frac{-1}{2 k_B T} \int_{0}^{1} \frac{1}{x^2} \frac{d}{dx} \left ( x \frac{dE_{i,j}}{dx} \right)  \exp \left [  \frac{-1}{k_B T}  \left (   \frac{x}{2}  \frac{dE_{i,j}}{dx}  - E_{i,j}   \right )  \right ]     dx.
\end{equation} 
Finally the correction factor which interpolates between the continuum and kinetic regime is given by \citet{Alam1987}
\begin{equation}
V_{i,j} = \frac{W^c_{i,j}[1+ 4 (D_{m,i} +D_{m,j} )/ \sqrt{\overline{v_i}^2+\overline{v_j}^2} (a_i+a_j)  ]}{1 + (W^c_{i,j}/W^k_{i,j})(4(D_{m,i} +D_{m,j})/\sqrt{\overline{v_i}^2+\overline{v_j}^2} (a_i+a_j) ) }.
\end{equation}

\begin{figure}[ht]
   \centering
\includegraphics[width=13cm]{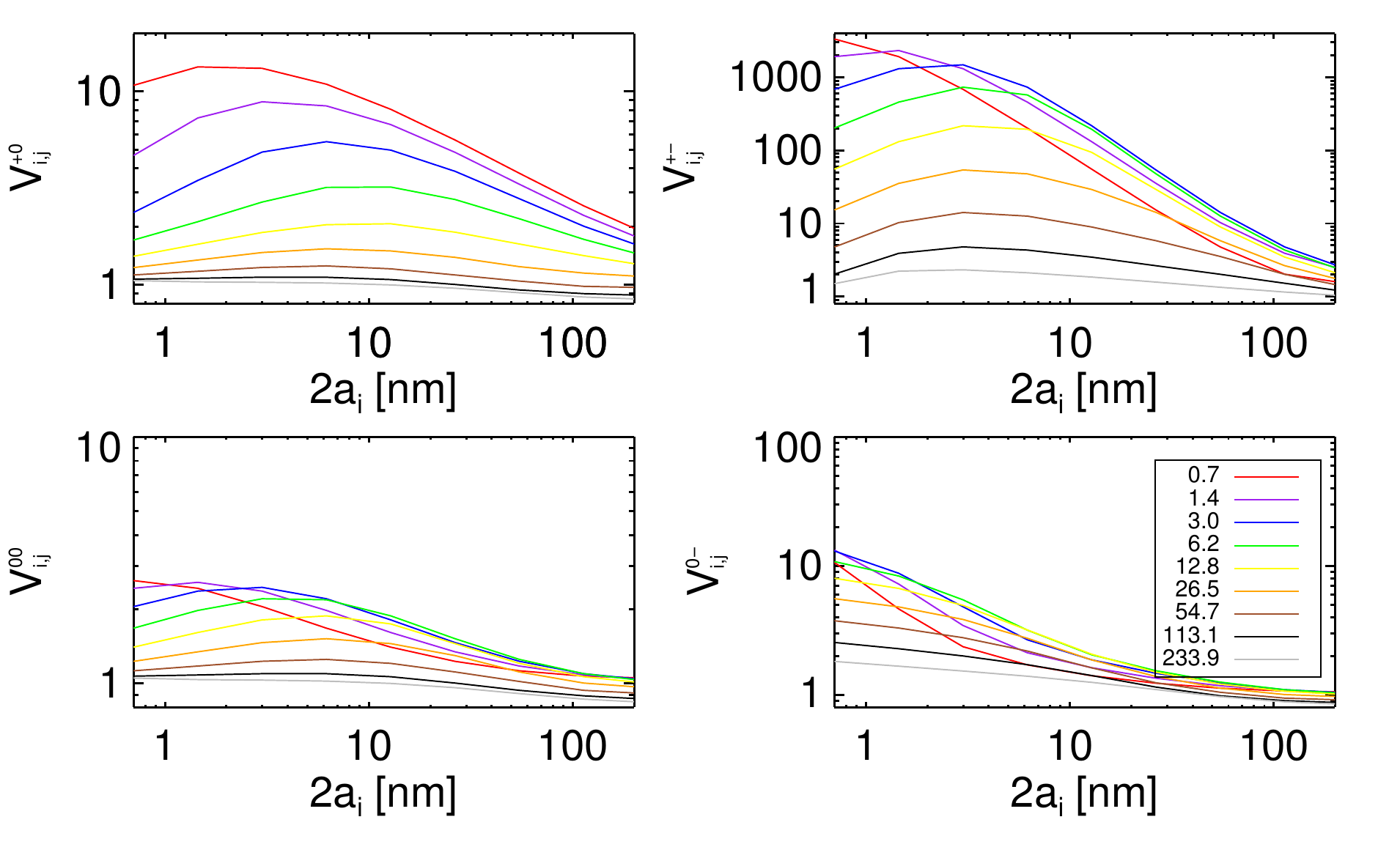}
   \caption{Enhancement factors between a particle of size $a_i$ and $a_j$ due to all effects mentioned in the Appendix and for $T=300\,$K, $\eta=1.8362\times10^{-5}\,\mathrm{kg}\,\mathrm{m}^{-1}\,\mathrm{s}^{-1}$ and mean free path $\lambda = 65\,\mathrm{nm}$, for different charges on each of the coagulating spheres. Numbers in the legend are diameters are diameters in units of nm.}
   \label{fig:V}
\end{figure}

Figure \ref{fig:V} graphically displays the calculated enhancement factors.
The above Brownian coagulation kernel can be used as a basis for including electrostatic and Van der Waals forces. Such effects are conveniently included by multiplying the Brownian coagulation kernel by a correction or enhancement factor. Such that 
\begin{eqnarray}
\kappa_{i,j}^{+0}   &=&  {\beta^B}_{i,j} V_{i,j}^{+0},  \nonumber \\
\kappa_{i,j}^{00}   &=&  {\beta^B}_{i,j} V_{i,j}^{00}, \nonumber \\
\kappa_{i,j}^{0-}    &=&  {\beta^B}_{i,j} V_{i,j}^{0-}. \nonumber \\
\end{eqnarray}
The above calculations do not include three-body interactions such as discussed in \citet{xerxes2013}, where charged aerosols are slowed down by collisions with (but not attachments to) a third body, and subsequently attaches to another aerosol of a given charge. This  will be important in the recombination between small ion-pairs, and that is the likely reason that the $V_{i,j}^{+-}$ coefficients for the smallest particles does not reach the experimentally determined recombination coefficient $\alpha = 1.6 \cdot 10^{-6}$ cm$^3$s$^{-1}$ but is a factor $\approx$ 5 too low. It could be corrected by simple adjustment, which we implement as 
\begin{equation}
\kappa_{i,j}^{+-} = \frac{{\beta^B}_{i,j} V_{i,j}^{+-}}
{1+\left(  \frac{ {\beta^B}_{i,j} V_{i,j}^{+-}}{\alpha}   -1\right)\exp{\left(-\frac{\sqrt{d_i^2+d_j^2}}{d_s}\right)}}, \nonumber
\end{equation}
where $d_s=100\,$nm is a characteristic scale length. The condensation coefficients used in the condensation equations are
\begin{eqnarray}
\beta_k^{00} &=& {\beta^B}_{0,k} V_{0,k}^{00},  \nonumber \\
\beta_k^{0+} &=& {\beta^B}_{0,k} V_{0,k}^{0+}, \nonumber\\
\beta_k^{0-}  &=& {\beta^B}_{0,k} V_{0,k}^{0-}. \nonumber\\
\beta_k^{+0} &=& {\beta^B}_{+,k} V_{+,k}^{+0}. \nonumber\\
\beta_k^{-0}  &=& {\beta^B}_{-,k} V_{-,k}^{-0}. \nonumber\\
\end{eqnarray}
Note that these $\beta$ only have one index $k$ for the radius of the aerosol $a_k$, as the radius of the condensing monomers (denoted with subscripts $0$,$-$, and $+$) are fixed at radii $a_0$, $a_-,$ and $a_+$. Again the recombination coefficient is used as a limit for the final two terms 
\begin{eqnarray}
\beta_k^{-+}  &=& \frac{{\beta^B}_{-,k} V_{i,j}^{-+}}{1+\left(  \frac{ {\beta^B}_{-,k} V_{-,k}^{-+}}{\alpha} -1\right)\exp{\left(-\frac{a_j}{a_s}\right)}}, \nonumber \\ 
\beta_k^{+-}  &=& \frac{{\beta^B}_{+,k} V_{i,j}^{+-}}{1+\left(  \frac{ {\beta^B}_{+,k} V_{+,k}^{+-}}{\alpha} -1\right)\exp{\left(-\frac{a_j}{a_s}\right)}}, \nonumber
\end{eqnarray}
where $\alpha = 1.6 \cdot 10^{-6}$ cm$^3$s$^{-1}$.

\bibliographystyle{naturemag}
\bibliography{bibliography}

\acknowledgments
JS is funded by the Danish council for independent research under the project “Fundamentals of Dark Matter Structures”, DFF - 6108-00470. NJS wishes to thank the generous support of the Israeli Ministry of Energy, grant no 218-11-036. All aerosol simulation data in this work is produced using the ION-CAGE v0.9 code descibed in the main text, and interaction coefficients described in the appendix. Both will be available on a Zenodo repository upon article acceptance, and output reproduced using parameters stated in the main text.
\end{document}